\documentclass[12pt]{article}
\usepackage{graphicx,rotating,epsfig,amsmath,amssymb,color,multirow}
\newlength{\digitwidth} 

\newcommand{\bra}[1]{\langle #1|}
\newcommand{\ket}[1]{|#1\rangle}
\newcommand{\braket}[2]{\langle #1|#2\rangle}
\newcommand{\Pro}{{\sf P}}
\newcommand{\Qz}{{\bf Q}}
\newcommand{\qj}{{\bf q}_j}
\newcommand{\tr}{{\rm tr}}
\newcommand{\Nj}{\{N_j\}}

\newcommand{\hpartj}{\{h_{n_j}\}}
\newcommand{\e}{{\rm e}}
\newcommand{\ee}{{e$^+$e$^-$}}
\newcommand{\ppb}{p$\bar{\rm p}$}
\newcommand{\gs}{\gamma_S}
\newcommand{\phiv}{\boldsymbol{\phi}}
\newcommand{\muv}{\boldsymbol{\mu}}
\newcommand{\tpm}{{$\pm$}}
\newcommand{\eecc}{$\e^+ \e^- \to {\rm c}\bar{\rm c}$}
\newcommand{\eebb}{$\e^+ \e^- \to {\rm b}\bar{\rm b}$}

\begin{document}
%%%%%%%%%%%%%%%%%%%%%%%%%%%%%%%%%%%%%%%%%%%%%%%%%%%%%%%%%%%%%%%%%%%%%%%%%%%%%%%%%%%%%%%%%
\begingroup

\thispagestyle{empty}
\baselineskip=14pt
\parskip 0pt plus 5pt

\begin{center}
{\large }
\end{center}

\bigskip
\begin{flushright}
%CERN--PH--EP\,/\,2008--019\\
%November 12, 2008
\end{flushright}

\bigskip
\begin{center}
{\LARGE\bf\boldmath
The QCD confinement transition:\\[0.3 cm]
hadron formation}

\bigskip\bigskip\bigskip

Francesco Becattini\\
Universit\`a di Firenze and INFN Sezione di Firenze, \\
Via G. Sansone 1, I-50019, Sesto Fiorentino (Firenze), Italy \\
{\it becattini@fi.infn.it}\\[0.5cm]
and \\[0.5 cm]
R. Fries\\
Cyclotron Institute, Texas A \& M University, \\
College Station, TX 77843, USA \\ 
and RIKEN/BNL Research Center, Brookhaven National Laboratory,
Upton, NY 11973, USA\\
{\it rjfries@comp.tamu.edu}

\bigskip\bigskip\bigskip

\end{center}

\begingroup
\leftskip=0.4cm
\rightskip=0.4cm
\parindent=0.pt

\bigskip

%Keywords: 
%PACS numbers: 

\endgroup
\bigskip\bigskip\bigskip

\vfill
%\begin{center}
%\emph{To be published in JHEP}
%\end{center}
%\HRule

\endgroup

~
%
% Questa istruzione se vuoi l'indice all'inizio
%
%\tableofcontents
%\newpage
%\cleardoublepage

\pagenumbering{arabic}
\setcounter{page}{1}

%%%%%%%%%%%%%%%%%%%%%%%%%%%%%%%%%%%%%%%%%%%%%%%%%%%%%%%%%%

%***********************************************************************
\section{Introduction}
%***********************************************************************

A major phenomenon that the theory of strong interactions, quantum chromodynamics 
(QCD), should account for is confinement: quarks and gluons are not observable 
particles. In fact, every physical process involving strong interactions at high 
energy results in the formation of hadrons, in which quarks and gluons are 
confined on a distance scale of ${\cal O}(1)$ fm. While, up to now, there is no formal 
proof that QCD implies confinement, there are many indications, both from perturbative
and from lattice numerical studies, that this is likely the case. Perturbative
QCD is applicable to scattering processes of quarks and gluons involving large 
momentum transfer ($\gg 1\;$ GeV) because the strong coupling constant $\alpha_S$ 
is small enough to allow a series expansion. However, this is no longer possible 
at a scale of 1 GeV or below, where the perturbative expansion is meaningless, and 
where confinement and hadronization, the process of hadron formation, takes place. 
Thus, hadronization is not yet calculable from QCD first principles and one has
to resort to phenomenological models. While this may seem an inconvenient limitation, 
still much can be learned from these models about QCD in the confinement regime. 
Indeed, if they are able to effectively describe the essential features of the 
actual physical process, they give us relevant information about the characteristics
of the fundamental theory. 

In this chapter, we will review two of these models, that have found widespread
use in relativistic heavy ion collisions. The first is a model with a rather long
history that has recently been revived by its successes in the description of hadronic 
multiplicities, the statistical model. This model is applicable to hadronization
in elementary collisions as well as heavy ion collisions. This is in fact its main 
strength, i.e.\ it captures a universal feature in the hadronization process. 

The second model is the quark recombination or coalescence model which extends the 
concept of single parton fragmentation function, which has been used in 
elementary collisions since the '70s. Its recent success comes from observations 
specific to relativistic heavy ion collisions.

We will start by reviewing the foundations and the main results of the statistical
model in Sect.~2 and of the quark recombination model in Sect.~3. In Sect.~4 we will 
compare the two models and discuss further perspectives in the understanding of the
hadronization phenomenon.

%***********************************************************************
\section{The Statistical Hadronization Model}
%***********************************************************************

The idea of applying statistical concepts to the problem of multi-particle
production in high energy collisions dates back to a work of Fermi \cite{fermi} in
1950, who assumed that particles originated from an excited region evenly occupying 
all available phase space states. This was one of Fermi's favorite ideas and soon 
led to an intense effort in trying to work out the predictions of inclusive particle rates 
calculating, analytically and numerically, the involved multidimensional phase-space 
integrals. When it became clear that the (quasi) isotropic particle emission in 
the center-of-mass frame predicted by Fermi's model was ruled out by the data, an 
amendment was put forward by Hagedorn \cite{hage} in the '60s, who postulated the 
existence of two hadron emitting sources flying apart longitudinally in the 
center-of-mass frame of a pp collision. Thereby, one could explain the striking
difference between spectra in transverse and longitudinal momentum. Hagedorn was 
also able to explain the almost universal slope of $p_T$ spectra in his renowned 
statistical bootstrap model, assuming that resonances are made of hadrons and 
resonances in turn.

After QCD turned up, many phenomenological models of strong interactions were no 
longer pursued and the statistical model was no exception. The resurgence of 
interest in these ideas came about when it was argued that a completely equilibrated
hadron gas would be a clear signature of the formation of a transient Quark-Gluon
Plasma (QGP) in heavy ion collisions at high energy. While it has been indeed 
confirmed that an (almost) fully equilibrated hadron gas has been produced 
\cite{various} in those collisions, the interest in this model was also 
revived by the unexpected observation that it is able to accurately reproduce 
particle multiplicities in elementary collisions \cite{elem}. Naively, one did
not expect a statistical approach to work in an environment where the number of
particles is ${\cal O}(10)$ because it was a belief of many that a hadronic
thermalization process would take a long time if driven by hadronic collisions.
Apparently this is not the case and one of the burning questions, which is
still waiting a generally accepted answer, is why a supposedly non-thermal
system exhibits a striking thermal behavior.
%----------------------------------------------------------------------------
\begin{figure}[htb]
\begin{center}
\includegraphics[scale=.3]{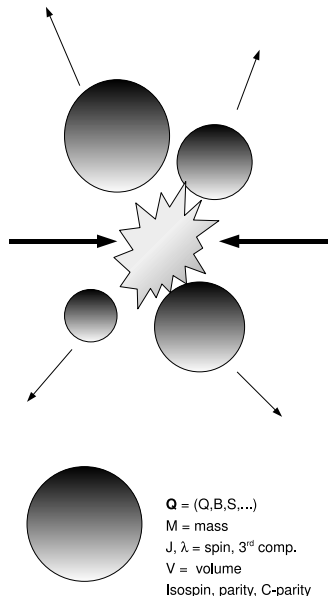}
\caption{High energy collisions are assumed to give rise to multiple clusters
at the hadronization stage [top]. Each cluster
[bottom] is a colorless extended massive object endowed with abelian charges
(electric, strange, baryonic etc.), intrinsic angular momentum and other quantum
numbers such as parity, $C$-parity and isospin.}
\label{collision}
\end{center}     
\end{figure}
%----------------------------------------------------------------------------

Before we address this interesting issue, it is appropriate to provide a rigorous
formulation of the model in a modern form, which is necessarily different from
Fermi's original model due to the tremendous improvement in our knowledge
of strong interactions phenomenology. The Statistical Hadronization Model (SHM) must 
be considered as an effective model describing the process of hadron formation in 
high energy collisions at energy (or distance) scales where perturbative QCD is no longer
applicable. A high energy collision is thought of as a complex dynamical process, 
governed by QCD, which eventually gives rise to the formation of extended massive 
colorless objects defined as {\em clusters} or {\em fireballs} (see Fig.~\ref{collision}). 
While the multiplicity, masses, momenta and charges of these objects are determined 
by this complex dynamical process, the SHM postulates that hadrons are formed from 
the decay of each cluster in a purely statistical fashion, that is:

\begin{quote}
{\em Every multihadronic state localized within the cluster and compatible with
conservation laws is equally likely}.
\end{quote}

This is the {\it urprinzip} of the SHM. The assumption of the eventual formation 
of massive colorless clusters is common for many hadronization models (e.g. the
cluster model implemented in the Monte-Carlo code HERWIG \cite{herwig}) based on the 
property of color preconfinement \cite{preconf} exhibited by perturbative QCD.
The distinctive feature of the SHM is that clusters have a finite spacial size. 
This aspect of clusters as a relativistic massive extended objects coincides with that 
of a bag in the MIT bag model \cite{bag}. Indeed, the SHM can be considered as an 
effective model to calculate bag decays. 

The requirement of finite spacial extension is crucial. If the SHM is to be an effective 
representation of the QCD-driven dynamical hadronization process, this characteristic 
must be ultimately related to the QCD fundamental scale $\Lambda_\mathrm{QCD}$. As we will 
see, the universal soft scale shows up in the approximately constant energy density at 
hadronization; in other words, the volume of clusters is in a constant ratio with 
their mass when hadronization takes place. It is also worth stressing here that there 
is clear, independent evidence of the finite size of hadronic sources in high 
energy collisions. Quantum interference effects in the production of identical 
particles, the so-called Bose-Einstein correlations or Hanbury Brown-Twiss second-order
interference, is by now a firmly established phenomenon. This effect would simply be 
impossible without a finite volume.

%********************************************************************
\subsection{Localized States}
\label{local}
%********************************************************************

The basic postulate of the Statistical Hadronization Model asserts that every 
localized multihadronic state which is contained within a cluster and is compatible 
with conservation laws is equally likely. The word {\em localized}, implying a finite 
spacial size, plays a crucial role, as we have emphasized. Thus, before getting to 
the heart of the SHM formalism, it is necessary to pause and clarify the distinction 
between localized and asymptotic states.

Such a difference is not an issue when the volume is sufficiently larger than 
the Compton wavelength of hadrons and it is disregarded in most applications where 
clusters supposedly meet this requirement (e.g. heavy ion collisions); yet, it is an 
important point at a fundamental level. Although in thermodynamics
the focus is on the limit of infinite volumes, we must start from a finite
volume and localized systems to introduce concepts like energy density, temperature
etc. Furthermore, in the hadronic world, finite size effects must be diligently 
taken into account when the volume is comparable to the (third power of) the pion's
Compton wavelength, $\sim 1.4$ fm. 

%----------------------------------------------------------------------------
\begin{figure}[htb]
\includegraphics[scale=.4]{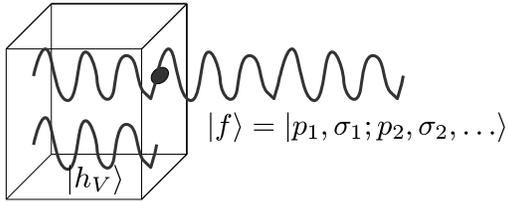}
\caption{The localized multi-hadronic states $\ket{h_V}$ pertaining to the quantum
field problem in a limited region. Asymptotic states $\ket{f}$ are the usual free
states characterized by particle momenta and spin components.}
\label{loca}     
\end{figure}
%----------------------------------------------------------------------------

The difference between a localized and an asymptotic state is depicted in Fig.~\ref{loca}.
For a single particle in a Non-Relativistic Quantum Mechanics (NRQM) framework, 
the conceptual difference is easier to grasp: a localized state is described by a 
wavefunction which vanishes outside the cluster's region whereas an asymptotic state is a 
wavefunction which is defined over the whole space (e.g. a plane or a spherical wave).
In Quantum Field Theory (QFT), a localized state is a state of the Hilbert space defined 
by the localized problem, e.g. the problem of the quantum field in the cluster's 
finite region. For a free field, if we enforce fixed or periodic boundary conditions, 
such states are simply defined by integer occupation numbers for each allowed mode in 
the finite region, as is well known. For a multiparticle state of non-interacting 
hadrons, this state will be defined by all occupation numbers of the modes determined
by the fields associated with the different species of hadrons, and we will simply denote 
it with $\ket{h_V}$ (where $h$ stands for ``hadrons'' and $V$ stands for the finite region 
of volume $V$). On the other hand, an asymptotic state is a state of the Hilbert 
space defined by the quantum field operators over the whole space; for a free field, 
these are the familiar multi-particle free states defined by, e.g., momentum and polarization:
\begin{equation}
\ket{f} = \ket{p_1,\sigma_1,\ldots,p_N,\sigma_N}
\end{equation}
There is a noteworthy and deep difference between non-relativistic quantum mechanical 
and quantum field theoretical case. In the latter, particle number is not fixed
and a localized state with multiplicity (defined as the sum of all occupation numbers) $N$ 
does not necessarily correspond to an asymptotic state with particle multiplicity 
$N$\footnote{We note that the multiplicity of an asymptotic free state is the
properly defined number of particles}. Unlike in NRQM, a localized state with multiplicity 
$N$ has non-vanishing projections over asymptotic states with different particle 
multiplicities. In symbols:
\begin{equation}
 \ket{N}_V = \alpha_{0,N}\ket{0}+\alpha_{1,N}\ket{1}+\ldots+
 \alpha_{N,N} \ket{N}+\ldots
\end{equation}
with the obvious condition that $\alpha_{i,N} \to 0$ when $V \to \infty$ for $i \ne N$.
In particular, the vacuum of the finite-region problem $\ket{0}_V$ is different from
the vacuum of the full-space problem $\ket{0}$, which is commonly known as the Casimir 
effect. With a straightforward mapping of the Hilbert space of the localized
quantum field onto the full Hilbert space, it is possible to express destruction 
operators of the localized field as linear combinations of destruction {\em and} 
creation operators of the field defined over the whole space \cite{microfield1}. 
These relations would be sufficient to calculate the coefficients of the above
equation, but this is not really needed in the SHM, as it will be soon clear.

%********************************************************************
\subsection{The Formalism: Basics}
\label{formalism}
%********************************************************************

Let us consider a cluster and assume first that it can be described as a {\em mixture} 
of states. Then, the basic postulate implies that the corresponding density 
matrix is a sum over all localized states projected onto the initial cluster's 
quantum numbers:
\begin{equation}\label{densmat}
\hat \rho \propto \sum_{h_V} \Pro_i \ket{h_V} \bra{h_V} \Pro_i 
\equiv \Pro_i \Pro_V \Pro_i
\end{equation}
where $\ket{h_V}$ are multi-hadronic localized states and $\Pro_i$ is the projector 
onto the cluster's initial conserved quantities: energy-momentum, intrinsic angular 
momentum and its third component, parity and the generators of inner symmetries
of strong interactions \footnote{Operators in the Hilbert space will be denoted with
a hat. Exceptions to this rule are projectors, which will be written in serif font,
i.e. $\Pro$.}. 

The operator $\Pro_i$ can be formally defined 
as the projector onto an irreducible vector of the full symmetry group and worked 
out in a group theory framework \cite{micro1,meaning}. It can be factorized into a 
"kinematic" projector, associated to general space-time symmetry, and a projector
for inner symmetries. For the space-time symmetry, the relevant group is the  
extended orthochronous Poincar\'e group IO(1,3)$^\uparrow$ and an irreducible state
is defined by a four-momentum $P$, a spin $J$ and its third component $\lambda$
and a discrete parity quantum number $\pi = \pm 1$. Therefore:
\begin{equation}\label{factor1}
   \Pro_i = \Pro_{P,J,\lambda,\pi} \Pro_{inner}  
\end{equation}
If the projector $\Pro_{P,J,\lambda}$ is worked out in the cluster's rest frame
where $P=(M,{\bf 0})$, it further factorizes into the product of simpler projectors
\cite{micro1,microfield2}, i.e.:
\begin{equation}\label{factor2} 
 \Pro_{P,J,\lambda,\pi}= \delta^4(P - \hat P) \Pro_{J,\lambda} 
 \frac{{\sf I} + \pi \hat \Pi }{2} 
\end{equation}
where $\hat P$ is the four-momentum operator, $\Pro_{J,\lambda}$ is a projector 
onto SU(2) irreducible states $\ket{J,\lambda}$ and $\hat{\Pi}$ is the space
reflection operator. 

As clusters are color singlets by definition, the projector $\Pro_{inner}$
involves flavor and baryon number conservation. 
In principle, the largest symmetry group one should consider is SU(3), plus three 
other U(1) groups for baryon number, charm and beauty conservation. However, SU(3)
symmetry is badly broken by the mass difference between strange and up, down quarks, 
so it is customary to take a reduced SU(2)$\otimes$U(1) where SU(2) is associated 
with isospin and U(1) with strangeness. The isospin SU(2) symmetry is
explicitly broken as well, but the breaking term is small and can generally be neglected.
However, most calculations in the past have replaced isospin SU(2) with another
U(1) group for electric charge, so that the symmetry scheme, from an original 
SU(2)$_{isospin} \otimes$U(1)$_{strangeness}\otimes$U(1)$_{baryon}$ reduces to 
U(1)$_{charge}\otimes$U(1)$_{strangeness} \otimes$U(1)$_{baryon}$.

Altogether, $\Pro_{inner}$ can be written as
\begin{equation}
 \Pro_{inner} = {\Pro}_{I,I_3} {\Pro}_\Qz {\Pro}_\chi 
\end{equation}
where $I$ and $I_3$ are isospin and its third component, $\Qz = (Q_1,\ldots,Q_M)$ 
is a vector of $M$ integer abelian charges (baryon number, strangeness, etc.)
and $\Pro_\chi$ is the projector onto C-parity, which makes sense only if the system 
is completely neutral, i.e. $I=0$ and $\Qz = {\bf 0}$; in this case, ${\sf P}_\chi$ 
commutes with all other projectors. 
 
From the density matrix (\ref{densmat}) the probability of observing an asymptotic 
multiparticle state $\ket{f}$ is
\begin{equation}\label{prob}
 p_f \propto \bra{f} \Pro_i \Pro_V \Pro_i \ket{f} 
\end{equation}
which is well-defined in terms of positivity and conservation 
laws. In fact, (\ref{prob}) is manifestly positive definite and $p_f = 0$ if 
the state $\ket{f}$ has not the same quantum numbers as the initial state.
By summing over all states $\ket{f}$, one obtains the trace of the operator $\Pro_i 
\Pro_V \Pro_i$ which is
\begin{equation}\label{sumprob}
 \sum_f p_f \propto \tr (\Pro_i \Pro_V \Pro_i) = \tr (\Pro^2_i \Pro_V) = a 
 \tr (\Pro_i \Pro_V) \, .
\end{equation}
The constant $a$ is divergent and positive. It can be directly checked by 
choosing the $\ket{f}$ as momentum eigenstates and using the 
expression on the right hand side of (\ref{factor2}). The reason for its presence
is the non-compactness of the Poincar\'e group, which makes
it impossible to have a properly normalized projector. The last trace in 
(\ref{sumprob}) can be written as
\begin{equation}
 \tr (\Pro_i \Pro_V) = \sum_{h_V} \bra{h_V} \Pro_i \ket{h_V} \equiv \Omega
\end{equation}
which is, by definition the {\em microcanonical partition function} \cite{microfield1}, 
i.e. the sum over all localized states projected onto the conserved quantities
defined by the selected initial state. If only energy and momentum conservation is 
enforced, $\Omega$ takes on a more familiar form:
\begin{equation}
 \Omega = \sum_{h_V} \bra{h_V} \delta^4(P-\hat P) \ket{h_V}   \, .
\end{equation}

Although the mixture of states defined by Eq.~(\ref{densmat}) allows us to 
calculate probabilities of any measurement unambiguously, a cluster could in 
principle also be described by a pure quantum state. Actually, the mixture of states 
only expresses our ignorance about the true state of the system, which is in principle
a pure one, or, more precisely, a pure state entangled with pure states of other 
clusters. To avoid slipping into fundamental quantum mechanics problems of decoherence 
and measurement, we take a pragmatic stance here. It suffices to realize 
that in some low-energy collision events, only one cluster might be created whose 
state is then necessarily a pure one. According to the postulate of the SHM, this must
be an even superposition of all localized states with the initial conserved 
quantities, i.e.\
\begin{equation}\label{pure}  
\ket{\psi} = \sum_{h_V} c_{h_V} \Pro_i \ket{h_V}  \qquad {\rm with}\,\, |c_{h_V}|^2=
 {\rm const}  \, .
\end{equation}
The probability of observing a final state $\ket{f}$ is then
\begin{eqnarray}\label{probpure}
&&|\braket{f}{\psi}|^2 = | \sum_{h_V} \bra{f}\Pro_i \ket{h_V} c_{h_V} |^2 \\ \nonumber
= && {\rm const} \sum_{h_V} |\bra{f}\Pro_i \ket{h_V}|^2
+\sum_{h_V \ne h'_V} \bra{f}\Pro_i \ket{h_V} \bra{h'_V} \Pro_i \ket{f} c_{h_V} c^*_{h'_V}
\, .
\end{eqnarray}
If the coefficients $c_{h_V}$ have random phases, the last term in Eq.~(\ref{probpure}) 
vanishes and we are left with the same expression appearing in Eq.~(\ref{prob}); 
in other words an effective mixture description is recovered.
Hence a new hypothesis is introduced in the SHM here: if the cluster is a pure state, 
the superposition of multi-hadronic localized states must have random phases.

Now the main goal of the model is to determine the probabilities (\ref{prob}) which 
involves the calculation of the projector $\Pro_V = \sum_{h_V} \ket{h_V}\bra{h_V}$,
a more limited task than the explicit calculation of all scalar products 
$\braket{h_V}{f}$. Since the states $\ket{h_V}$ are a complete set of states of 
the Hilbert space $H_V$ for the localized problem, the above projector is simply
a resolution of the identity of the localized problem and can be written in the
basis of the field states. For a real scalar field this is
\begin{equation}\label{proje}
   \Pro_V = \int_V \mathcal{D} \psi \ket{\psi}\bra{\psi}
\end{equation}
where $\ket{\psi} \equiv \otimes_{{\bf x}} \ket{\psi({\bf x})}$ and $\mathcal{D}\psi$ 
is the functional measure; the index $V$ means that the functional integration must 
be performed over the field degrees of freedom in the region $V$, that is 
${\mathcal D} \psi = \prod_{{\bf x}\in V} d \psi({\bf x})$. One has to face several 
conceptual subtleties in the endeavor of calculating the probabilities (\ref{prob})
with the projector (\ref{proje}), e.g.\ how to deal with field boundary 
conditions and with their values outside the region $V$. However, by enforcing the known
non-relativistic limit is possible to come to an unambiguous and consistent result
\cite{microfield1}.

%********************************************************************
\subsection{Rates of Multiparticle Channels}
\label{rates}
%********************************************************************

According to the formulae introduced in the previous section, the decay rate of 
a cluster into a {\em channel} $\Nj$ ($\Nj$ is the array of multiplicities 
$(N_1,\ldots,N_K)$ for hadron species $1,\ldots,K$) is proportional 
to the right hand side of (\ref{prob}) integrated over momenta and summed over
polarization states of the final hadrons. Taking into account only energy-momentum 
conservation, so that the projector (\ref{factor2}) reduces to
$$
 \Pro_{PJ\lambda\pi} \to \Pro_{P} = \delta^4(P-\hat P) \, ,
$$
and neglecting quantum statistics effects, this is proportional to the microcanonical 
partition function with fixed particle multiplicities \cite{micro1,microfield1}
\begin{equation}\label{boltz}
 \Omega_{\Nj} = \frac{V^N}{(2\pi)^{3N}} \left( \prod_{j=1}^K 
 \frac{(2S_j+1)^{N_j}}{N_j!} \right)
 \int d^3 p_1 \ldots \int d^3 p_N \; \delta^4(P_0 - \sum_i p_i)
 \bra{0} \Pro_V \ket{0} \, .
\end{equation}
Here $N$ is the number of particles, $S_j$ the spin, $P_0=(M,{\bf 0})$, $M$ is 
the mass and $V$ is the cluster's proper volume. This formula is the same as it
would be obtained in NRQM, with the factor $\bra{0} \Pro_V \ket{0}$ 
(which becomes 1 in the limit $V \to \infty$) being the only effect of the field 
theoretical treatment \cite{microfield1}. Since only relative rates make sense, 
this common factor for all channels is irrelevant.

Loosely speaking, Eq.~(\ref{boltz}) tells us that the decay rate of a
massive cluster into some multi-hadronic channel is proportional to its phase 
space volume. However, it should be emphasized that the "phase space volume" in 
(\ref{boltz}) is calculated with the measure $d^3 x \, d^3 p$ for each particle, 
and not with the one usually understood in QFT, i.e. $d^3 p/2\varepsilon$. Although 
this is also commonly known as "phase space", it is quantitatively different from 
the properly called phase space measure $d^3 x \, d^3 p$ and should be called 
"invariant momentum space" measure \cite{chai}.

Eq.~(\ref{boltz}) can be cast in a form which makes its Lorentz invariance apparent. 
Define a four-volume $\Upsilon = V u $ \cite{chai} where $V$ is the
cluster's rest frame and $u$ its four-velocity vector. Then (\ref{boltz}) can
be rewritten as:
\begin{eqnarray}\label{boltz2}
 \Omega_{\Nj} = && \frac{1}{(2\pi)^{3N}} \left( \prod_{j=1}^K 
 \frac{(2S_j+1)^{N_j}}{N_j!} \right)
 \int d^4 p_1 \ldots \int d^4 p_N \; \\ \nonumber 
 && \left[\prod_{i=1}^N \Upsilon \cdot p_i 
 \delta (p^2_i-m^2_i) \theta(p^0_i) \right] \delta^4(P_0 - \sum_i p_i)
 \bra{0} \Pro_V \ket{0}
\end{eqnarray}
which is manifestly covariant. In this form it can be directly compared with the 
general formula for the decay rate of a massive particle into a $N$-body channel:
\begin{equation}\label{decay}
 \Gamma_N \propto \sum_{\sigma_1,\ldots,\sigma_N} \frac{1}{(2\pi)^{3N}} 
 \left( \prod_j \frac{1}{N_j!} \right) \int \frac{d^3 p_1}{2\varepsilon_1} 
 \ldots \int \frac{d^3 p_N}{2\varepsilon_N} \; 
 |M_{fi}|^2 \delta^4(P_0 - \sum_i p_i)
\end{equation}
where $\sigma$ labels, as usual, polarization states. Comparing (\ref{boltz}) 
with (\ref{decay}) we can infer a dynamical matrix element for the SHM which 
is
\begin{equation}
  |M_{fi}|^2 \propto \prod_{i=1}^N \Upsilon \cdot p_i \, .
\end{equation}
Therefore, according to the SHM the dynamics in cluster decay is limited to a 
common factor for each emitted particle, which linearly depends on the cluster's 
spacial size. The four-volume $\Upsilon$ is simply proportional to the four-momentum 
of the cluster through the inverse of energy density $\rho$ and therefore:
\begin{equation}
  |M_{fi}|^2 \propto \frac{1}{\rho^N} \prod_{i=1}^N P \cdot p_i \, .
\end{equation}
This expression explicitly shows the separation between the kinematic arguments
of the dynamical matrix element, and the scale $1/\rho$ which determines particle
production. This ought to be ultimately related to the fundamental scale of quantum
chromodynamics, $\Lambda_\mathrm{QCD}$.

It has already been stressed that the finite cluster size is the distinctive
feature of the SHM. This peculiarity of the model stands out when taking into account 
quantum statistics for the calculation of decay rates. Our final result, 
for which (\ref{boltz}) is a special case when all particles belong to different 
species, reads
\begin{eqnarray}\label{quantum}
 \Omega_{\Nj} &=& \int d^3 {\rm p}_1 \ldots d^3 {\rm p}_N  \; 
  \delta^4 (P_0-\sum_{i=1}^N p_i) \prod_j \sum_{\hpartj} 
 \frac{(\mp 1)^{N_j + H_j} (2S_j + 1)^{H_j}}
{\prod_{n_j=1}^{N_j} n_j^{h_{n_j}} h_{n_j}!} \\ \nonumber 
 &\times& \prod_{l_j=1}^{H_j} F_{n_{l_j}} \; \bra{0} \Pro_V \ket{0}
\end{eqnarray}
where $\hpartj$ is a {\em partition} of the integer $N_j$ in the multiplicity 
representation, that is $\sum_{n_j=1}^{N_j}  n_j h_{n_j} = N_j$,   
$\sum_{n_j=1}^{N_j} h_{n_j} = H_j$ and $\sum_j N_j = N$. The factor $F_{n_{l_j}}$
in Eq.~(\ref{quantum}) are Fourier integrals:
\begin{equation}\label{fourier}
  F_{n_l} = \prod_{i_l=1}^{n_l} \frac{1}{(2\pi)^3} \int_V d^3 {\rm x} \; 
 \e^{i {\bf x \cdot}({\bf p}_{c_l(i_l)}-{\bf p}_{i_l})}
\end{equation}
over the cluster's region $V$, $c_l$ being a cyclic permutation of order $n_{l}$.
The expression (\ref{quantum}) has been obtained in refs.~\cite{micro1,microfield1} 
and is a generalization of a similar one calculated by Chaichian, Hagedorn
and Hayashi \cite{chai} whose validity is restricted to large volumes. It is a 
so-called {\em cluster decomposition} of the microcanonical partition function of 
the channel. For sufficiently large volumes, all terms in Eq.~(\ref{quantum}) turn 
out to be proportional to the $H_j$th power of the volume $V$ \cite{micro1}, so 
that the leading term is the one with $H_j = N_j$ for all $j$, which leads precisely 
to Eq.~(\ref{boltz}). Thus, Eq.~(\ref{quantum}) is a generalization of (\ref{boltz})
containing all corrections due to quantum statistics.

In general, with respect to the Boltzmann case (\ref{boltz}), the channel rate is 
enhanced by the presence of identical bosons and suppressed by that of fermions. 
This means that Bose-Einstein and Fermi-Dirac correlations are built into  
the SHM. The reader has probably anticipated this fact through the appearance of typical 
Fourier integrals in the cluster decomposition. This feature of the SHM is an
almost obvious consequence of the cluster's finite spacial size.

%*********************************************************
\subsection{Interactions}
\label{interactions}
%*********************************************************

So far, we have dealt with non-interacting particles. However, the localized hadronic
fields we have used to calculate transition probabilities do interact and this must 
be taken into account. The energy of the interacting system must be conserved until 
the final asymptotic multi-hadronic state is reached which is made of particles stable 
under strong interactions, namely pions, kaons, nucleons and octet hyperons. 

Formally, this implies that the projector (\ref{factor2}) must include the interacting 
Hamiltonian in the $\delta^4(P-\hat P)$ operator\footnote{In all virtually known 
field theories, there is no additional interacting term for momentum and angular 
momentum \cite{weinberg}.}. The definition (\ref{prob}) for the probability 
to observe an asymptotic state $\ket{f}$, has to be modified by the insertion of 
M\o ller operator $\hat \Omega$, yet summing over the complete set of states yields 
the same result as in Eq.~(\ref{sumprob}):  
\begin{equation}\label{traceint}
 \sum_f p_f \propto \tr (\Pro_i \Pro_V \Pro_i) = a \, \tr (\Pro_i \Pro_V)
 = a \, \sum_{h_V} \bra{h_V} \Pro_i \ket{h_V} \equiv a\, \Omega
\end{equation}
where $a$ is an irrelevant divergent constant and $\Omega$ is the microcanonical 
partition function of the interacting hadronic system. 

An outstanding theorem by Dashen, Ma and Bernstein (DMB) \cite{dmb} allows us to 
calculate the microcanonical partition function of an interacting system in the
thermodynamic limit $V \to \infty$ as the sum of the free one plus a term depending 
only on the physical scattering matrix. It can be expressed as
\begin{equation}\label{dashen}
\tr \delta^4(P-\hat P) = \tr \delta^4 (P- \hat P_0) + 
 \frac{1}{4\pi i} \tr \left[ \delta^4 (P- \hat P_0) 
 \hat{{\cal S}}^{-1}\frac{\stackrel{\leftrightarrow}
 {\partial}}{\partial E}\hat{{\cal S}}\right]
\end{equation}
where $\hat P$ includes the full interaction Hamiltonian, whereas $\hat P_0$ 
only contains the free one; ${\cal S}$ is the reduced scattering matrix on the 
energy-momentum shell.
If more conserved quantities other than energy and momentum are involved, like 
those encountered in section \ref{formalism}, the theorem is readily extended
and relevant projectors can be placed next to the $\delta$-functions in 
Eq.~(\ref{dashen});
it suffices that these conserved quantities are associated with symmetries of 
both free and interacting theory.

%-------------------------------------------------------------------------- 
\begin{figure}[htb]
\begin{minipage}[t]{55mm}
\includegraphics[width=13pc]{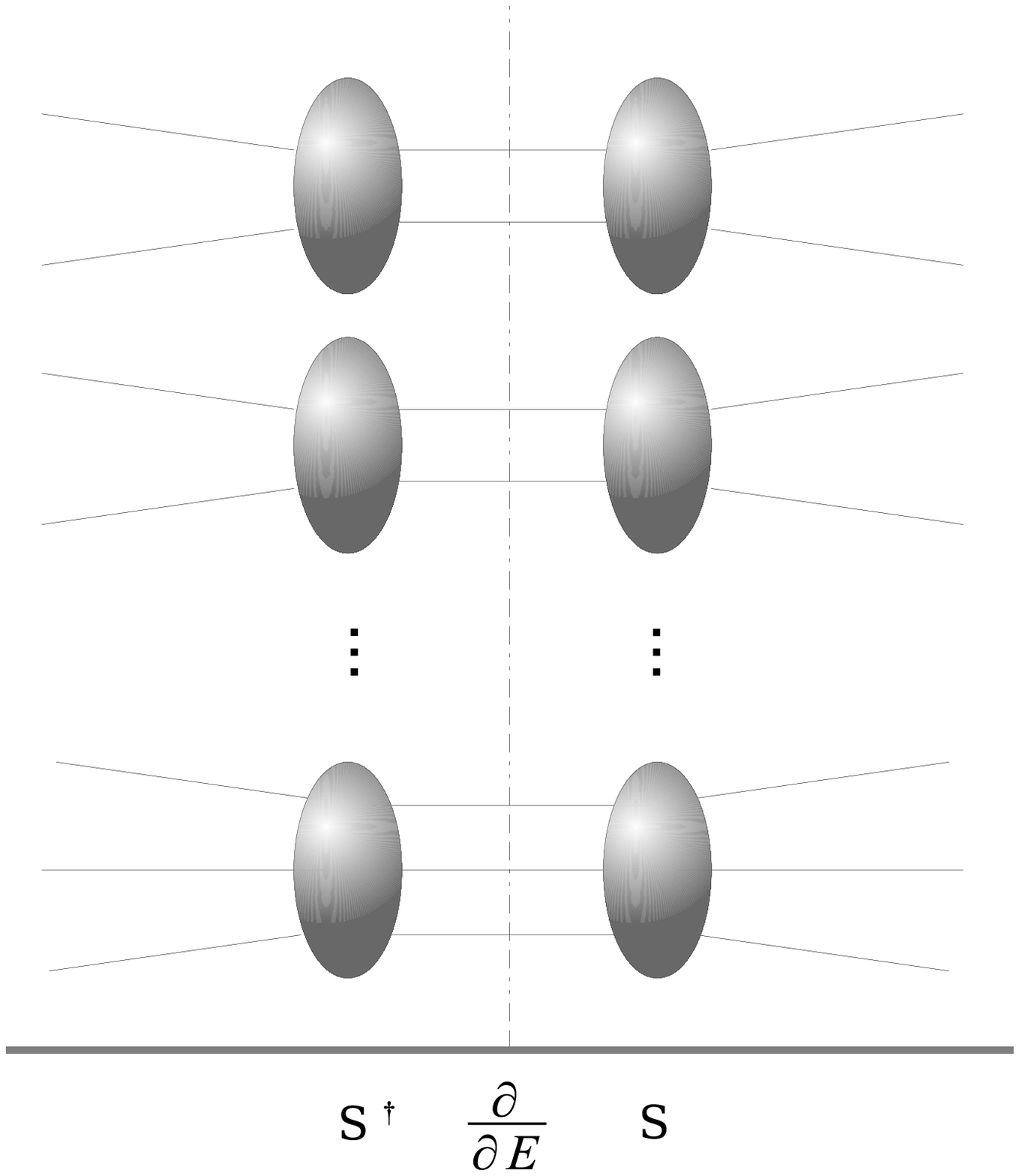}
\end{minipage}
%
%\hspace{\fill}
%
\begin{minipage}[t]{55mm}
\includegraphics[width=13pc]{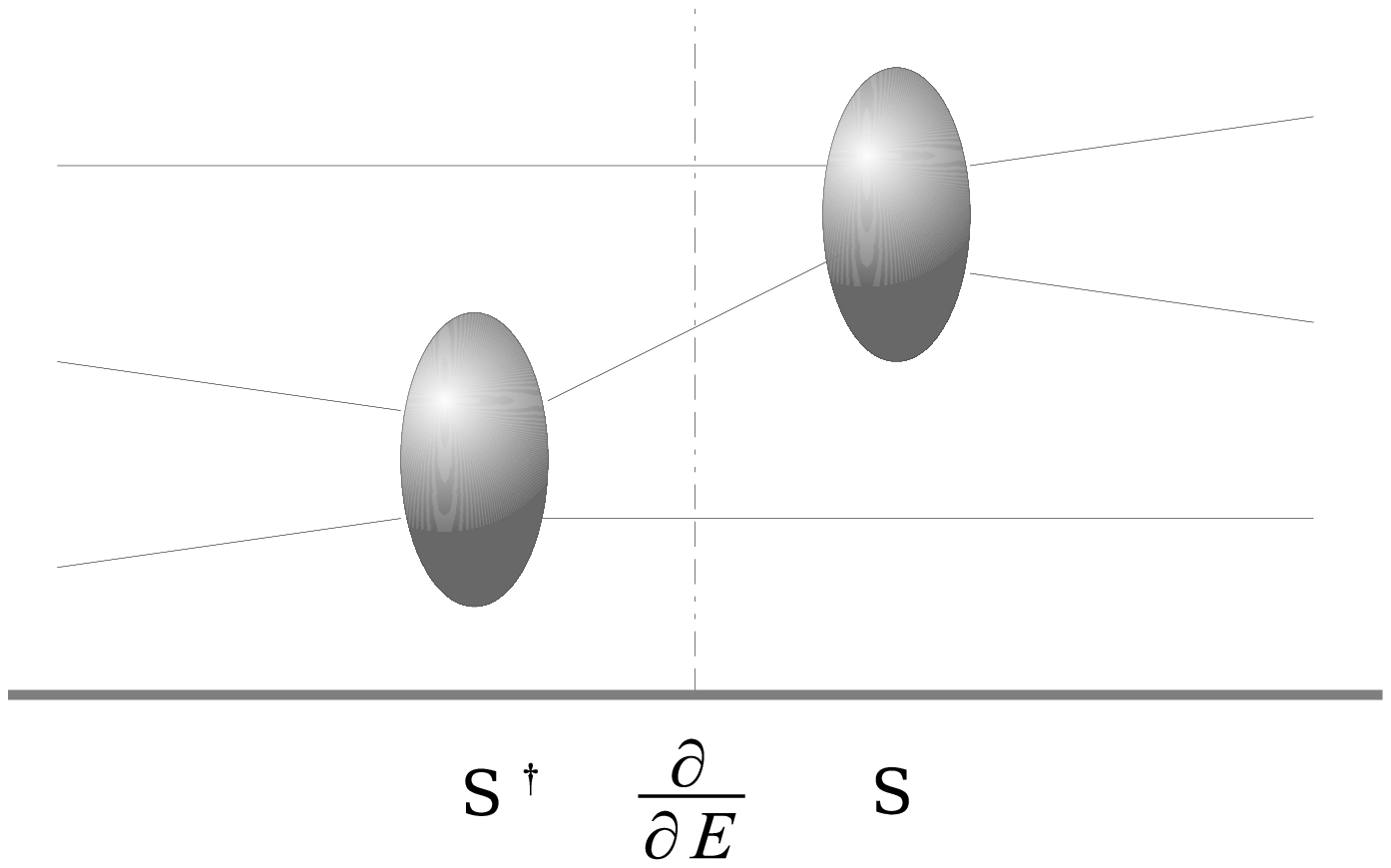}
\end{minipage}
\caption{Left panel:symmetric diagrams for the cluster decomposition of the 
interaction term in the DMB theorem. Right panel: non-symmetric diagrams.}
\label{dmb}
\end{figure}
%---------------------------------------------------------------------------

This theorem is indeed the starting point of the {\em hadron-resonance gas} model
since it can be shown that if only the resonant part of the scattering matrix
is retained and the background interaction can be neglected, the main contribution
of the second term on the right hand side of Eq.~(\ref{dashen}) is equivalent 
to considering all hadronic resonances as free particles with distributed mass.
More specifically, if a cluster decomposition of the two scattering operators is 
carried out in Eq.~(\ref{dashen}), the corresponding diagrams can be divided in 
two sets: the symmetric diagrams (see Fig.~\ref{dmb}, left panel) and the non-symmetric
ones (see Fig.~\ref{dmb}, right panel). Taking into account that the terminal legs 
on both sides have to be the same stable particles on entry and exit (we are 
calculating a trace), it can be shown that the main contribution to symmetric 
diagrams comes from the matching resonances in bubbles facing each other.
For each term of the trace, this amounts to adding the decay products of resonances 
considered as free particles with masses distributed according to a relativistic
Breit-Wigner form. In symmetric diagrams, there is in principle an additional 
contribution from resonance interference, which might be non-negligible in case of wide,
overlapping resonances with the same decay channel, but it depends on mostly 
unknown complex parameters and it is thus neglected.

Likewise, the asymmetric diagrams give an additional contribution which also depends
on the aforementioned complex interference parameters. While the number of such 
diagrams greatly exceeds the symmetric ones due to the large number of resonances, 
contributing terms can be both positive and negative and hopefully a partial 
cancellation occurs when summing them up for a selected final state.

Altogether, retaining only the resonant interaction and symmetric diagrams in the
scattering matrix cluster decomposition, and neglecting resonance interference
leads to the following picture: an interacting hadron gas is, to a good approximation,
a gas of non-interacting free hadrons and resonances. Since non-resonant interaction
should be negligible, the ideal hadron-resonance gas picture holds if the energy density 
or temperature of the system is large enough for most resonances to be excited.
A quantitative assessment of how large these parameters are is still missing, a
rough estimate being $T > 100$ MeV.

An important remark is now in order. The DMB theorem affirms the equality of two 
traces, but not of single trace terms. Yet, the decomposition of Eq.~(\ref{dashen}), 
implying the ideal hadron-resonance gas picture, is widely used for the calculation 
of inclusive stable hadronic multiplicities as well, which requires a condition 
stronger than the equality of the traces on both sides. In other words, using the 
decomposition (\ref{dashen}) to calculate average multiplicities or fluctuations 
requires the equality to hold for multiparticle generating functions and not only 
for microcanonical partition functions. Up to now, the extension of (\ref{dashen}) 
to generating functions has never been proved; most likely, it does not hold 
and corrections to this assumption are necessary. Moreover, while the theorem 
requires the thermodynamic limit, it is commonly used at finite volume. These 
limitations should be always kept in mind when using the ideal hadron-resonance 
gas model.

%*********************************************************
\subsection{High Energy Collisions}
\label{collisions}
%*********************************************************
 
As we have seen in Sect.~\ref{rates}, each individual cluster produced in a high 
energy collision (shown in Fig.~\ref{collision}), should be hadronized according 
to formula (\ref{quantum}), or its approximation (\ref{boltz}), which yields 
the rates of a given channel within the microcanonical ensemble, including energy-momentum 
conservation. If clusters are large enough, the microcanonical ensemble could be
well approximated by the canonical \cite{micro1,micro2} or even grand-canonical 
ensemble for average multiplicities. This is not the case in elementary collisions 
(\ee, pp, etc.) while it is generally possible in heavy ion collisions, as we will
see later in this section. Calculating observables in high energy collisions within 
the SHM then implies summing microcanonical averages over all produced clusters
and this requires in turn knowledge of their charges and four-momenta. In fact, 
this latter information is unknown to the SHM and only a dynamical model of the 
pre-hadronization stage of the process (such as, e.g., HERWIG) can provide it. 

However, if we are interested in calculating Lorentz-invariant observables (such 
as average multiplicities) the momenta of clusters are immaterial and only charges 
and masses matter. In this case, one can introduce a peculiar extra-assumption 
which allows to considerably simplify the calculation. Basically, it is assumed 
that the probability distribution:
$$
  w({\bf Q_1},M_1;\ldots,{\bf Q}_N,M_N)
$$
of masses $M$ and conserved abelian charges ${\bf Q}$ for $N$ different clusters 
is the same as one would have by randomly splitting a large cluster (defined as 
{\em Equivalent Global Cluster}, EGC) into $N$ subsystems with given volumes. 
Thereby, the Lorentz invariant observables can be calculated for one (equivalent
global) cluster, whose volume is the sum of proper cluster volumes and whose charge 
is the sum of cluster charges, hence the conserved charge of the initial colliding 
system. The full mathematical procedure is described in detail in ref.~\cite{becagp}.

In such a global averaging process, the EGC generally turns out to be large enough 
in mass and volume so that the canonical ensemble becomes a good approximation of 
the more fundamental microcanonical ensemble \cite{micro2}; in other words, a 
temperature can be introduced which replaces the {\it a priori} more fundamental 
description in terms of energy density. It was shown that the mass of the EGC 
should be at least 8 GeV (with an energy density of 0.5 GeV/fm$^3$) for the canonical 
ensemble to be a reasonably good approximation \cite{micro2}. Also, it should be emphasized 
that in such a mathematical reduction process, temperature has essentially a global 
meaning and not local as in hydrodynamical models (see next subsection). The only 
meaningful local quantity in actual physical process are energy densities and 
individual physical clusters cannot be described in terms of a temperature, 
unless they are sufficiently large. Nevertheless, this ``global'' temperature 
closely mirrors the value of energy density at which clusters hadronize. Indeed,
it is this latter value which mainly determines hadronization-related observables;
the requirement of the charge distribution of EGC is a side-assumption which 
is important to simplify calculations, but it can possibly be replaced by other 
distributions leaving final results essentially unchanged.
 
In this approach, the primary multiplicity of each hadron species $j$ is given 
by \cite{becagp}:
\begin{equation}\label{mult}
 \langle n_j \rangle^{\rm primary} = \frac{V T (2S_j+1)}{2\pi^2} 
 \sum_{n=1}^\infty \gs^{N_s n}(\mp 1)^{n+1}\;\frac{m_j^2}{n}\;
 {\rm K}_2\left(\frac{n m_j}{T}\right)\, \frac{Z(\Qz-n\qj)}{Z(\Qz)}
\end{equation}
where $V$ is the (mean) volume and $T$ the temperature of the equivalent global 
cluster. Here $Z(\Qz)$ is the canonical partition function depending on the initial 
abelian charges $\Qz = (Q,N,S,C,B)$, i.e., electric charge, baryon number, 
strangeness, charm and beauty, respectively; $m_j$ and $S_j$ are the mass 
and the spin of the hadron $j$, $\qj = (Q_j,N_j,S_j,C_j,B_j)$ its corresponding 
charges; the upper sign applies to bosons and the lower sign to fermions. 

The parameter $\gs$ in (\ref{mult}) is an extra phenomenological factor implementing 
an {\it ad hoc} suppression of hadrons with $N_s$ strange valence quarks with 
respect to the equilibrium value. This parameter is outside a pure thermodynamical
framework and it is needed to reproduce the data, as we will see. For temperature 
values of 160 MeV or higher, Boltzmann statistics, corresponding to 
the term $n=1$ only in the series (\ref{mult}), is a very good approximation 
(within 1.5\%) for all hadrons but pions. For resonances, the formula (\ref{mult}) is 
folded with a relativistic Breit-Wigner distribution for the mass $m_j$. The final
multiplicities, to be compared with the data, are determined by adding to the 
primary multiplicity (\ref{mult}) the contribution from the decay of unstable 
heavier hadrons, according to the formula
\begin{equation}\label{branching}
\langle n_j \rangle  = \langle n_i \rangle^{\mathrm{primary}} + 
\sum_k \mathrm{Br}(k\rightarrow j) \langle n_k \rangle  \; .
\end{equation}
%
%----------------------------------------------------------------------------
\begin{figure}[htb]
\begin{center}
\includegraphics[scale=.45]{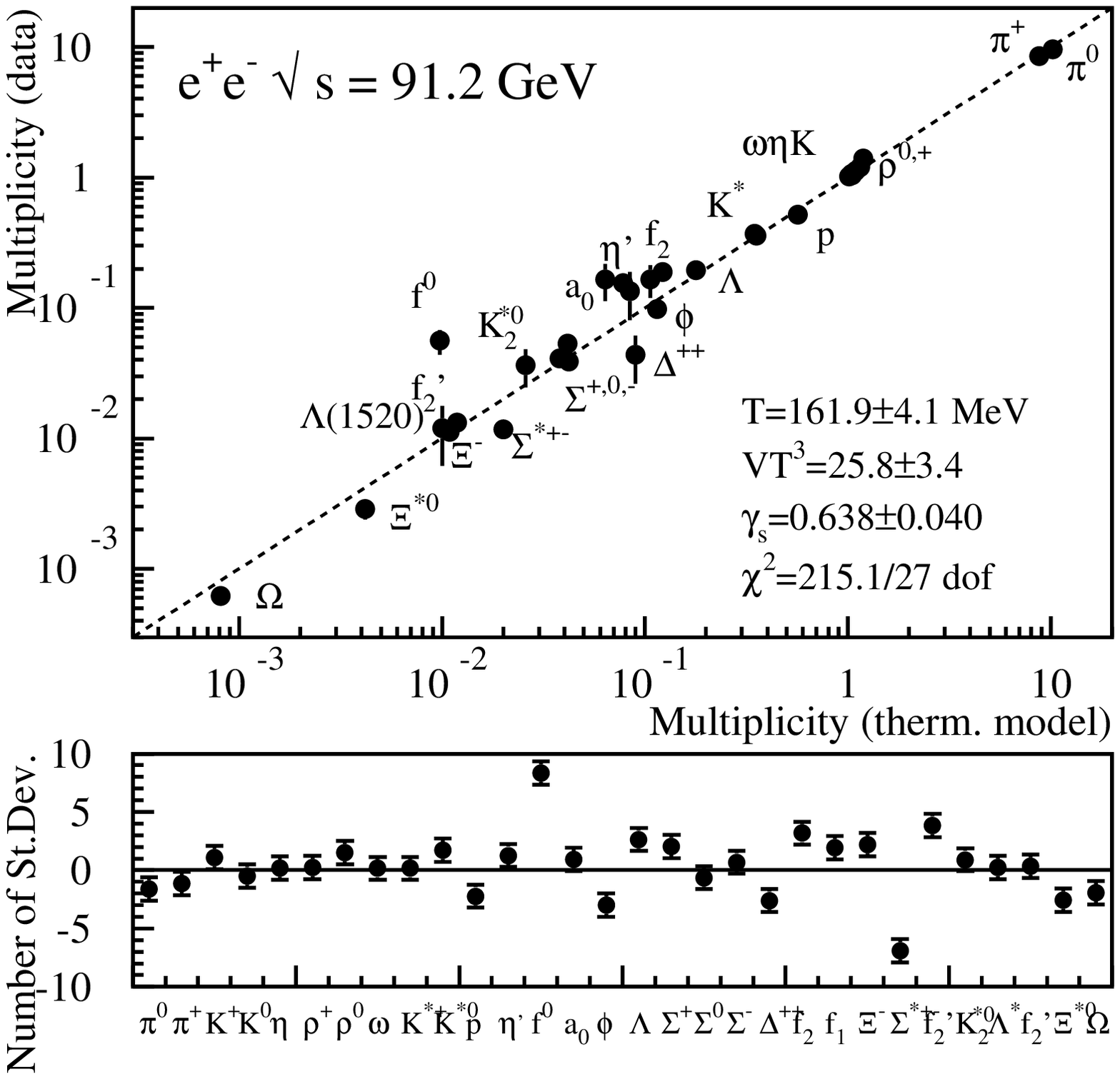}
\caption{Upper panel: measured vs theoretical multiplicities of light-flavoured
hadrons in \ee collisions at $\sqrt s = 91.25$ GeV. Lower panel: fit residuals
(from ref.~\cite{bcms}).}
\label{ee91} 
\end{center}    
\end{figure}
%----------------------------------------------------------------------------

The canonical partition function can be expressed as a multi-dimensional 
integral
\begin{eqnarray}\label{mpf}
 && Z(\Qz) = \frac{1}{(2\pi)^N} \int_{-\pi}^{+\pi} \, d^N \phi \,\, 
 \e^{i \Qz \cdot \phiv} \nonumber  \\ 
 && \times \exp \, \left[ \frac{V}{(2\pi)^3} \sum_j (2S_j+1) 
 \int d^3 p \,\, \log \, (1 \pm \gamma_s^{N_{sj}} 
  \e^{-\sqrt{p^2+m_j^2}/T_i -i \qj \cdot \phi})^{\pm 1} \right] 
\end{eqnarray}   
where $N$ is the number of conserved abelian charges. Unlike the grand-canonical 
case, the logarithm of the canonical partition function does not scale linearly 
with the volume. Therefore, the so-called {\em chemical factors} $Z(\Qz-n\qj)/Z(\Qz)$ 
\cite{antti} turn out to be less than unity even for a completely neutral system 
at finite volume (canonical suppression) and reach their grand-canonical value 1 
at asymptotically large volumes \cite{variouscf,elem}.
%----------------------------------------------------------------------------
\begin{figure}[htb]
\begin{center}
\includegraphics[scale=.45]{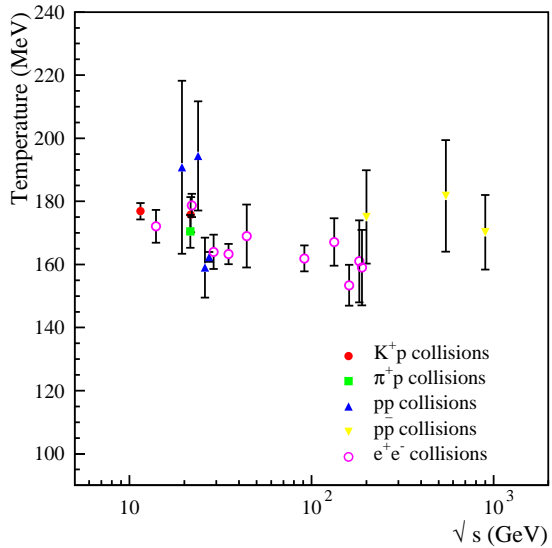}
\caption{Temperatures fitted in elementary collisions as a function of center-of-mass
energy.}
\label{temp}
\end{center}     
\end{figure}
%----------------------------------------------------------------------------
%----------------------------------------------------------------------------
\begin{table}[!h]
\begin{center}
\begin{tabular}{|c|c|c|c|c|}
\hline
   Particle  &        Experiment (E)   & Model (M)  & Residual & $(M - E)/E$  [\%] \\ 
\hline

$D^0$        & 0.559 \tpm 0.022   & 0.5406     & -0.83    & -3.2	  \\
$D^+$        & 0.238 \tpm 0.024   & 0.2235     & -0.60    & -6.1	  \\
$D^{*+}$     & 0.2377\tpm 0.0098  & 0.2279     & -1.00    & -4.1	  \\
$D^{*0}$     & 0.218 \tpm  0.071  & 0.2311     & 0.18	  &  6.0	  \\
$D^0_1$      & 0.0173 \tpm 0.0039 & 0.01830    & 0.26	  &  5.8	  \\
$D^{*0}_2$   & 0.0484 \tpm 0.0080 & 0.02489    & -2.94    & -48.6	  \\
$D_s$        & 0.116  \tpm 0.036  & 0.1162     &  0.006   &  0.19	  \\
$D^*_s$      & 0.069  \tpm 0.026  & 0.0674     & -0.06    & -2.4	  \\
$D_{s1}$     & 0.0106 \tpm 0.0025 & 0.00575    & -1.94    & -45.7	  \\
$D^*_{s2}$   & 0.0140 \tpm 0.0062 & 0.00778    & -1.00    & -44.5	  \\
$\Lambda_c$  & 0.079  \tpm 0.022  & 0.0966     & 0.80	  & 22.2	  \\
\hline \hline
$(B^0+B^+)/2$                	&  0.399 \tpm 0.011  &  0.3971    & -0.18   &  -0.49	 \\
$B_s$                        	&  0.098 \tpm 0.012  &  0.1084    &  0.87   &  10.6	 \\
$B^*/B$(uds)                 	&  0.749 \tpm 0.040  &  0.6943    & -1.37   &  -7.3	 \\
$B^{**}\times BR(B(^*)\pi)$  	&  0.180 \tpm 0.025  &  0.1319    & -1.92   &  -26.7	 \\
$(B^*_2+B_1)\times BR(B(^*)\pi)$&  0.090 \tpm 0.018  &  0.0800    & -0.57   &  -11.4	  \\
$B^*_{s2} \times BR(BK)$     	&  0.0093 \tpm 0.0024&  0.00631   & -1.24   &  -32.1	 \\
b-baryon                     	&  0.103 \tpm  0.018 &  0.09751   & -0.30   &  -5.3	 \\
$\Xi_b^-$                    	&  0.011 \tpm  0.006 &  0.00944   & -0.26   &  -14.2	 \\

\hline	
\end{tabular}
\bigskip
\caption{Abundances of charmed hadrons in \eecc~annihilations and bottomed
hadrons in \eebb~annihilations at $\sqrt s$ = 91.25 GeV, compared to the 
prediction of the statistical model (from ref.~\cite{bcms}).}
\label{hfstat}
\end{center}
\end{table}
%----------------------------------------------------------------------------

The light-flavoured multiplicities in \ee show a very good agreement with the 
predictions of the model, as it shown in Fig.~\ref{ee91}: the temperature value is 
about 160 MeV and the strangeness undersaturation parameter $\gs \sim 0.7$. Similar 
good agreements are found for many kinds of high energy elementary collisions over a large 
energy range \cite{elem}. Also, an excellent agreement between measured and predicted 
relative abundances of heavy flavoured hadronic species in \ee collisions by using 
the model parameters fitted to light-flavoured multiplicities \cite{elem,bcms},
as shown in Table~\ref{hfstat}. 
 
The overall striking feature is that the temperature turns out to be approximately constant 
over two orders of magnitude in centre-of-mass energy with a value of 160-170 MeV (see 
Fig.~\ref{temp}) and very close to the QCD critical temperature as determined from
lattice calculations. There must certainly be a profound connection between the 
thus-found hadronization temperature and QCD thermodynamics, a connection which has
not been made clear yet. Nevertheless, this finding indicates that hadronization 
is a universal process occurring at a critical value of the local energy density, i.e. 
when clusters have an energy density of $\simeq 0.5$ GeV/fm$^3$.

The parameter $\gs$ is found to be less than 1 in all examined elementary collisions,
ranging from $\sim 0.5$ in hadronic collisions to $\sim 0.7$ in \ee collisions (see
Fig.~\ref{gs}). This extra parameter most likely reflects the different mass of 
the strange quark with respect to lighter u, d quarks. This is a second scale, 
besides $\Lambda_{QCD}$, which must play a role in hadronization in view of its
value ${\cal O}(100)$ MeV. Altogether, one can say that the SHM description of 
hadronization is in excellent agreement with QCD at least with regard to the
the number of parameters. The two parameters $T$ and $\gs$ correspond to the
two fundamental scales $\Lambda_{QCD}$ and $m_s$, the strange quark mass. While 
we lack a definite relation connecting them (see, however ref.~\cite{bcms}), 
it is worth stressing that a phenomenological description of hadronization in 
terms of fewer parameters cannot be possible.

Finally, the statistical model shows a very good capability of reproducing 
transverse momentum spectra in hadronic \cite{becagp} as well as heavy ion collisions
\cite{heinzlectures}. Particularly, the phenomenon of approximate $m_T$ scaling 
observed in pp collisions is nicely accounted for by the model. However, the exact 
${\bf p}_T$ conservation at low energy and the increasing importance of jet emission 
at high energy restrict the validity of the statistical canonical formulae to a limited 
centre-of-mass energy range. Within this region, a clear consistency is found between 
the temperature parameter extracted from the spectra and that from average 
multiplicities. Altogether, this finding bears out one of the key predictions of the 
SHM, namely the existence of a definite relation between the dependence of particle 
production rates on mass and, for each particle species, their momentum spectra (in 
the cluster's rest frame) because they are both governed by one parameter, the 
energy density (or temperature) at hadronization.

%----------------------------------------------------------------------------
\begin{figure}[htb]
\begin{center}
\includegraphics[scale=.65]{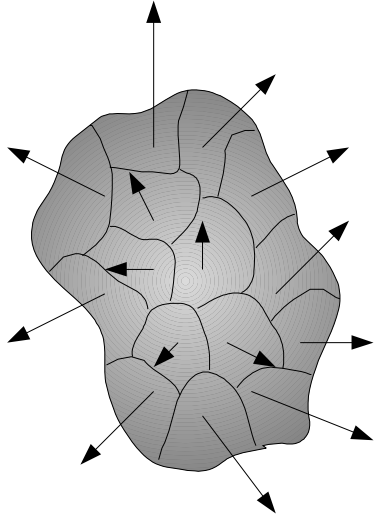}
\caption{Spacial distributions of clusters in heavy ion collisions according 
to the hydrodynamical picture. In this model, nearby clusters interact from
an early stage on and their momenta and charges are strongly correlated with
their positions, unlike in elementary collisions.}
\label{hi}  
\end{center}   
\end{figure}
%----------------------------------------------------------------------------
 
%*********************************************************
\subsection{Heavy Ion Collisions}
\label{heavy}
%*********************************************************

In heavy ion collisions, the system is much larger and two possibilities are 
usually envisaged: either hadronizing clusters are simply much larger than those 
in elementary collisions; or clusters are hydrodynamical cells, i.e. they are small 
but in thermal contact with each other due to previous thermalization, which implies 
a strong correlation between their position and momentum and charge densities 
(see Fig.~\ref{hi}).
In both case the canonical or grand-canonical formalisms apply to individual clusters.
For the former case, it is worth mentioning that the transition from a canonical to
a grand-canonical description effectively occurs when the cluster volume is of the 
order of 100 fm$^3$ at an energy density of 0.5 GeV/fm$^3$ \cite{antti}. If the 
EGC reduction assumption still applies, chemical factors in Eq.~(\ref{mult}) are 
replaced by fugacities, and in this case the phase-space integrated multiplicities 
read
\begin{equation}\label{multh}
 \langle n_j \rangle^{\rm primary} = \frac{V T (2S_j+1)}{2\pi^2} 
 \sum_{n=1}^\infty \gs^{N_s n}(\mp 1)^{n+1}\;\frac{m_j^2}{n}\;
 {\rm K}_2\left(\frac{n m_j}{T}\right)\, \exp[n \muv \cdot \qj/T] \, .
\end{equation}
$\muv$ is a vector of chemical potentials pertaining to the conserved abelian
charges, i.e. the electrical chemical potential $\mu_Q$, the baryon chemical potential
$\mu_B$ and the strangeness chemical potential $\mu_S$. Usually, but not always,
$\mu_S$ and $\mu_Q$ are determined by enforcing strangeness neutrality and by fixing
the ratio $Q/B$ to be the same as the initial $Z/A$ ratio of the colliding nuclei.

%----------------------------------------------------------------------------
\begin{figure}[htb]
\begin{center}
\includegraphics[scale=.45]{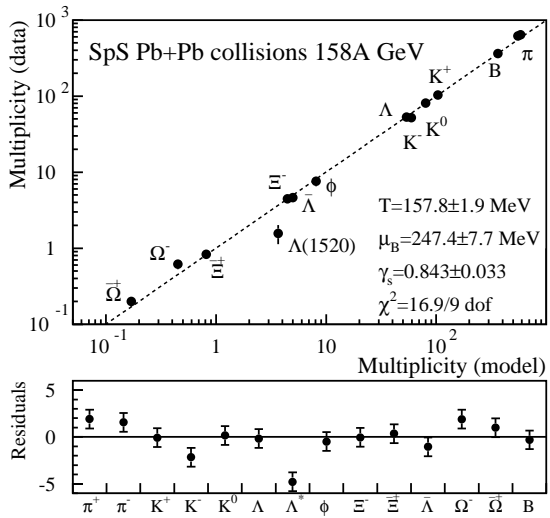}
\caption{Upper panel: measured vs theoretical multiplicities of light-flavoured
hadrons in Pb-Pb collisions at $\sqrt s_{NN} = 17.2$ GeV. Lower panel: fit residuals
(from ref.~\cite{becahi3}).}
\label{pbpb} 
\end{center}    
\end{figure}
%----------------------------------------------------------------------------

In the framework of the hydrodynamical model, formula (\ref{multh}) applies to
individual clusters identified with hydrodynamic cells and both temperature and
chemical potentials depend on space-time; when integrating particle densities
to get average multiplicities, one should take into account this dependence. It 
is important to stress that the hydrodynamical description is a salient feature 
of heavy ion collisions due to the early thermalization of the system in the partonic
phase, a phenomenon which does not occur in elementary collisions. It is this 
early thermalization which establishes the strong correlation between positions
and velocities of clusters, supposedly absent in elementary collisions. 
%----------------------------------------------------------------------------
\begin{figure}[htb]
\begin{center}
\includegraphics[scale=.85]{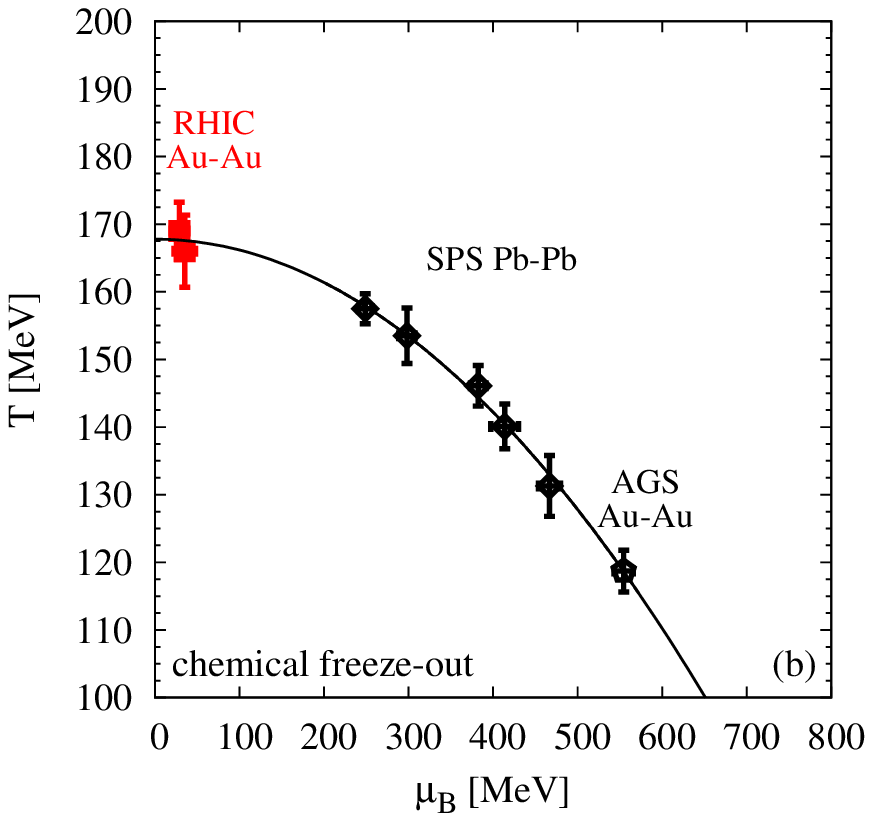}
\caption{Temperature vs baryon chemical potential fitted with multiplicities
in heavy ion collisions (from ref.~\cite{bm}).}
\label{tmu} 
\end{center}    
\end{figure}
%----------------------------------------------------------------------------
 
Provided that rapidity distributions are wide enough, and that there is little 
variation of the thermodynamical parameters of clusters around midrapidity, the 
formula (\ref{multh}) describes rapidity densities of hadrons at midrapidity as 
well: this condition is fulfilled at RHIC energies, but not at AGS and SPS energies, 
where the measured rapidity distributions are not significantly wider than those of a 
single fireball at the temperature found \cite{becahi4}. 

In general, the fits to particle multiplicities in heavy ion collisions are of 
the same good quality as in elementary collisions (see Fig.~\ref{pbpb}). Many
groups have analyzed the data over more than a decade \cite{various} and the overall 
description is very good throughout all explored energies and one finds a smooth curve 
in the $T-\mu_B$ plane (see Fig.~\ref{tmu}).

%*********************************************************
\subsection{Strangeness Production}
\label{strangeness}
%*********************************************************

The statistical model is a very useful tool to study one of the main features 
of relativistic heavy-ion collisions, the increase of relative strangeness
production with respect to elementary collisions. This was one of the early
signatures proposed for Quark-Gluon Plasma formation \cite{rafmul}, and it has
therefor attracted much attention both on the theoretical and experimental side.
According to the SHM, this is mainly an effect of the increase of the global volume 
from elementary to heavy-ion collisions. In elementary collisions, the EGC volume
is small enough for the chemical factors (see Eq.~(\ref{mpf})) of strange particles 
to be consistently less than 1 for systems with vanishing net strangeness, a 
phenomenon known as strangeness canonical suppression. 

\begin{figure}[htb]
\begin{minipage}[t]{55mm}
\includegraphics[width=14pc]{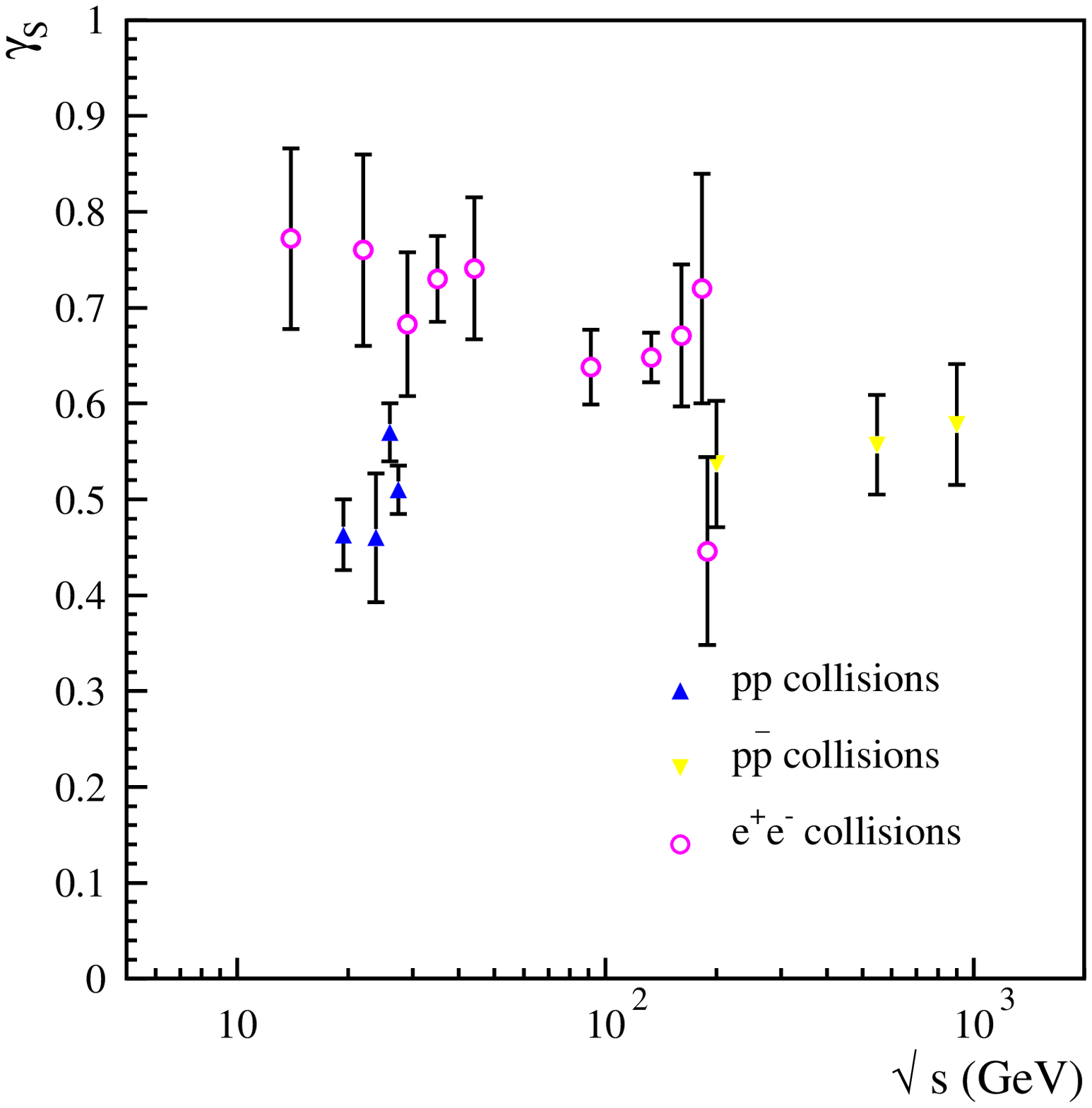}
\end{minipage}
%
%\hspace{\fill}
%
\begin{minipage}[t]{55mm}
\includegraphics[width=14pc]{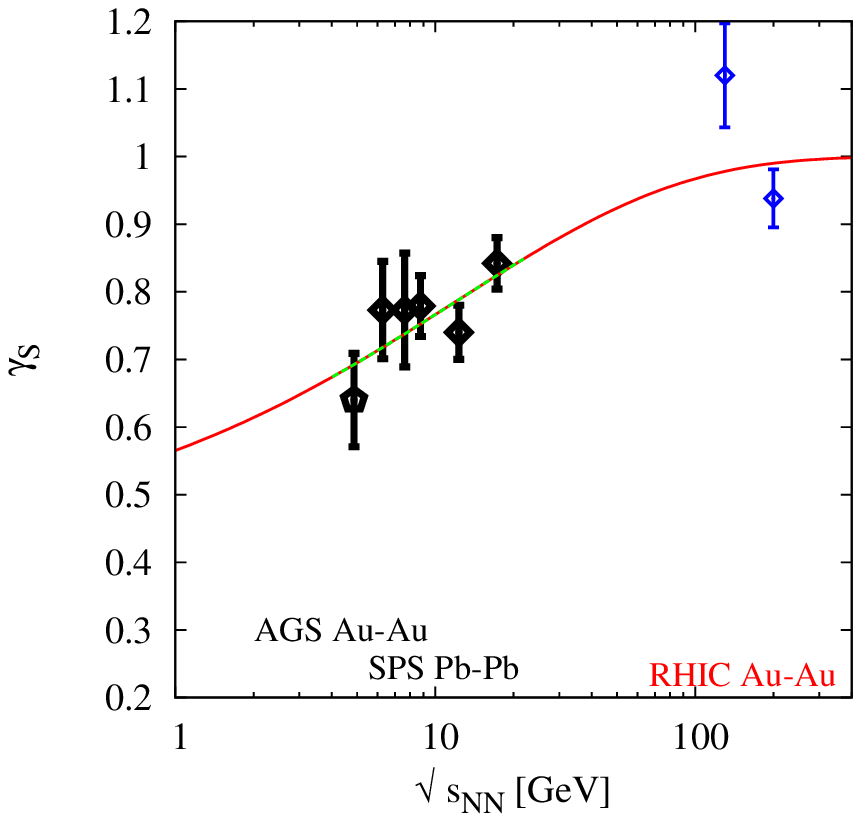}
\end{minipage}
\caption{Left panel: the strangeness undersaturation parameter $\gs$ as a 
function of center-of-mass energy in \ee, pp and \ppb\ collisions. Right
panel: the strangeness undersaturation parameter $\gs$ as a function of
center-of-mass energy in central heavy ion collisions (from ref.~\cite{bm}).}
\label{gs}
\end{figure}

However, canonical suppression is not enough to account for strangeness enhancement
from pp to heavy-ion collisions: also an increase of $\gs$ is needed. This is 
demonstrated by neutral mesons containing strange quarks, especially $\phi$ meson,
which do not suffer canonical suppression but are relatively more abundant
in heavy-ion collisions \cite{bgs,sollfrank}. Therefore, from a SHM viewpoint, one
can say that, as far as particle abundances is concerned, the only substantial 
difference between elementary and heavy ion collisions resides in the different 
$\gs$ values, which are generally higher in heavy-ion collisions and increase slowly 
as a function of center-of-mass energy (see Fig.~\ref{gs}): at RHIC 
energies one finds $\gs \simeq 1$ in central collisions. However, since $\gs$ is an 
empirical parameter which lies outside of a pure statistical mechanics framework, this 
observation does not clarify the origin of strangeness enhancement.

%-------------------------------------------------------------------------- 
%---------------------------------------------------------------------------

\begin{figure}[htb]
\begin{center}
\includegraphics[scale=.45]{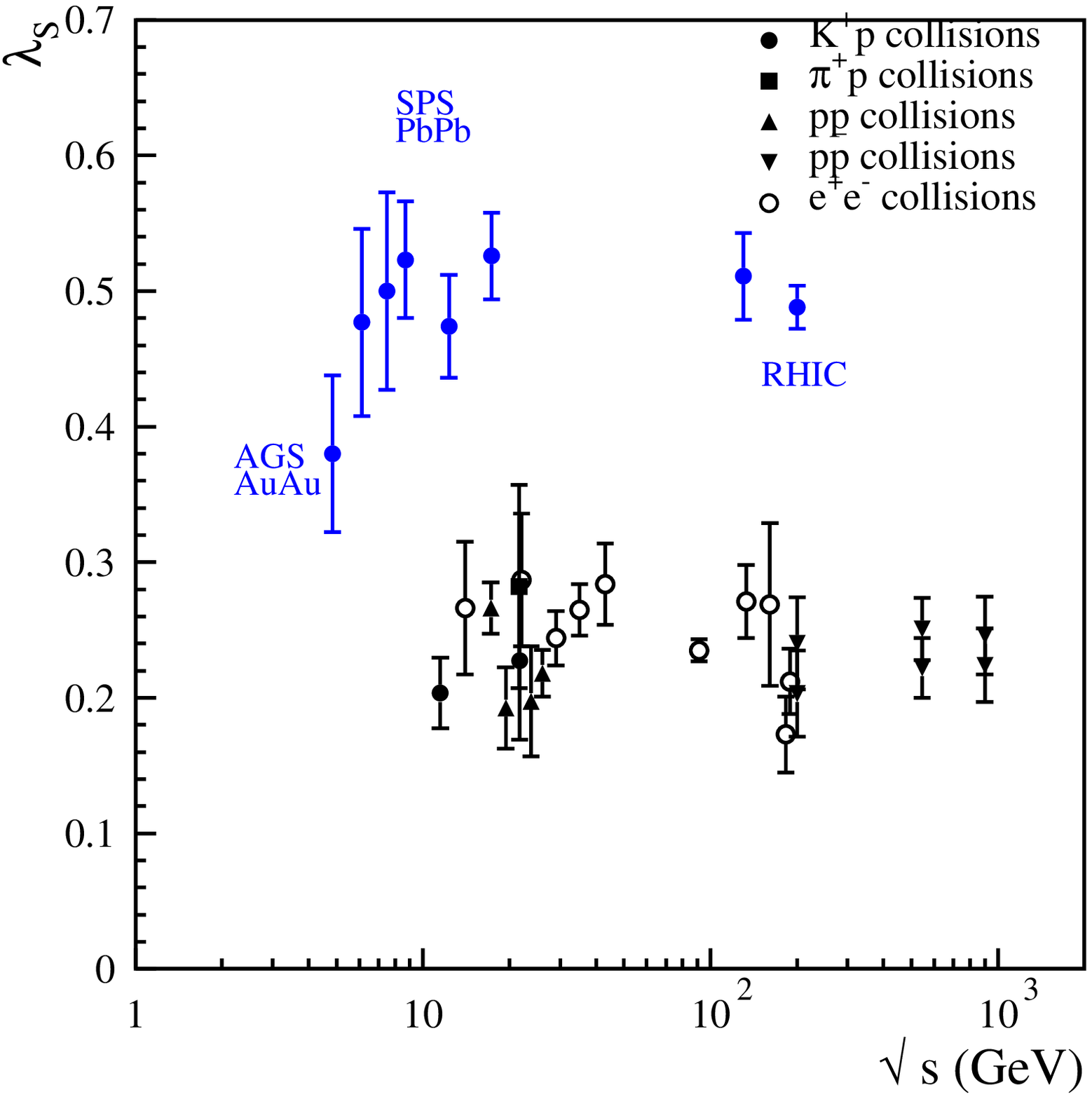}
\caption{Wroblewski ratio $\lambda_S$ (see text for definition) as determined in
elementary and heavy ion collisions from fits of multiplicities to the statistical 
hadronization model.}
\label{ls}   
\end{center}
\end{figure}

It is interesting to note that, while $\gs$ shows no special regularity in 
elementary collisions, the ratio of newly produced ${\bar{\rm s}}{\rm s}$ pairs 
over one half ${\bar{\rm u}}{\rm u}+{\bar{\rm d}}{\rm d}$ pairs (the so-called 
Wroblewski ratio $\lambda_S$) turns out to be around 0.2-0.25 at all energies, whereas 
it is definitely higher in heavy ion collisions (see Fig.~\ref{ls}).  

%----------------------------------------------------------------------------
%----------------------------------------------------------------------------

\begin{figure}[htb]
\begin{center}
\includegraphics[scale=.45]{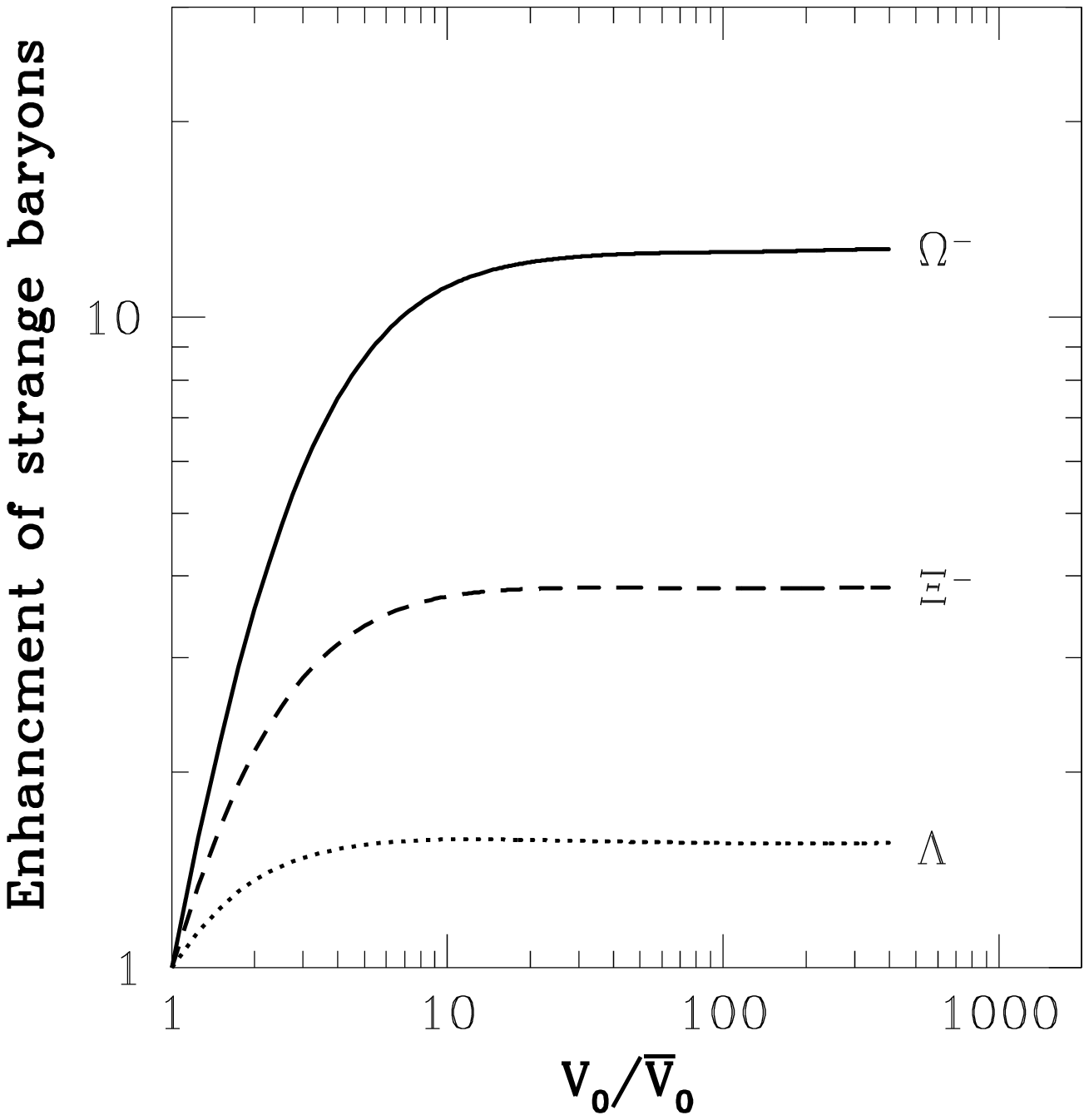}
\caption{Canonical enhancement (defined here as the ratio between the chemical 
factor $Z(S-S_j)/Z(S)$ and its value at some fixed volume $V_0$) as a function 
of volume for hyperons (from ref.~\cite{redlich}).}
\label{enhan}     
\end{center}
\end{figure}

These differences, both in $\gs$ and $\lambda_S$ have been and still are subject of 
investigation. Relevant information comes from the centrality dependence of
strangeness production, which provides an 
interpolation from single pp collisions at large values of nuclear impact parameter
to head-on heavy-ion collisions at low values thereof. The enhancement has been 
measured by the experiments WA97 and NA57 at SPS energy \cite{na57} and STAR at
RHIC \cite{starenhanc} for hyperons and other strange particles and it has been 
found to be hierarchical in strangeness content (highest for $\Omega^-$, lowest for
$\Lambda$). These observations led some authors \cite{redlich} to put forward a 
picture where $\gs$ is an effective parametrization of a canonical suppression. 
For large enough baryon number and charge, it is possible to take a mixed 
canonical-grand-canonical approach where only strangeness conservation is enforced, 
while electric and baryon-chemical potential are introduced. The chemical factors
$Z(S-S_j)/Z(S)$ depend on the volume and saturate at large volumes, as expected. Therefore, 
if we want to account for $\gs<1$ with this mechanism, there must be some 
small sub-regions within a large fireball where strangeness is exactly vanishing 
even for the most central collisions. Thereby, chemical
factors are significantly less than 1 and a suppression with respect to the 
grand-canonical limit is implied. However, this model has two major problems:
\begin{enumerate} 
\item{} Since measured enhancement steadily increases from peripheral to central 
collisions and hadronization temperature does not change \cite{star,cley,bm}, 
the volume of the sub-regions with $S=0$ should also increase and a saturation 
is thus expected (see Fig.~\ref{enhan}); yet, no saturation is observed,
which is quite an oddity. 
\item{} As has been mentioned, canonical suppression has no effect on $\phi$; 
yet, the relative yield of this meson with two constituent strange quarks is 
also observed to increase from peripheral to central collisions \cite{starphi}
and with $\gs=1$ and the observed constant temperature, this cannot occur.
\end{enumerate}

%----------------------------------------------------------------------------
%----------------------------------------------------------------------------

Recently, a geometrical explanation of these two features has been advocated 
\cite{bmcc} based on a superposition of emission from a hadron-resonance gas at 
full chemical equilibrium with $\gs=1$, defined as the {\it core}, and from 
nucleon-nucleon collisions at the boundary of the overlapping region of the two 
colliding nuclei, defined as the {\it corona}, from which produced particles escape 
unscathed. 
Since in NN collisions strangeness is suppressed with respect to a fully 
equilibrated, grand-canonical hadron gas, if such NN collisions account for a 
significant fraction of total particle production, a global fit to particle 
multiplicities will find $\gs<1$, as indeed observed in data. The idea of superposing
different sources is common to other models (a similar one is discussed in 
ref.~\cite{hohne}). The peculiar feature of this specific model is to assume 
single NN collisions as secondary sources; only in this case does it seem possible 
to reproduce centrality dependence of the $\phi$ meson.

%*********************************************************
\subsection{Thermalization: How Is It Achieved?}
%*********************************************************

After discussing the success of the SHM in reproducing particle multiplicities 
and the intriguing universality of its main parameter, the temperature, one is obviously
led to the question how this can come about. A classical process of thermalization 
through binary collisions between formed hadrons, advocated in heavy ion collisions 
\cite{pbm}, is ruled out in elementary collisions because the expansion rate is fast 
and hadrons are not interacting for a long enough time for this to take place. 
But even in heavy ion collisions peculiar features of the data cannot be explained 
in a hadronic kinetic picture \cite{kestin} without invoking the predominance of 
multi-body collisions; since, in this case, the hadronic mean free path is comparable
or smaller than Compton wavelengths, the collisional picture breaks down naturally.

There 
is evidence that thermalization occurs at a relatively early stage over a large 
region (i.e. clusters several femtometers wide) in heavy ion collisions, whereas it is a 
late phenomenon (i.e. very close to hadronization) occurring over small (of the order of 1 fm) 
distances in elementary collisions. Yet, the agreement between model and data is 
surprisingly accurate in elementary collisions, even more accurate than in 
heavy ion collisions, the only difference being in the level of strangeness phase space 
saturation. Somehow, the hadrons must {\em be born into equilibrium} as Hagedorn first 
pointed out \cite{hage2} and was reaffirmed by others \cite{others,stock}. 

The idea that this thermal-like behavior is of genuine quantum-mechanical origin and 
not related to semi-classical collision processes, is shared by many \cite{stock} and is
espoused in this review. A different point of view was presented in ref.~\cite{hsu} 
where it was argued that the thermal behavior could just be {\em mimicked} by a 
matrix element which is weakly dependent on the final kinematic variables in 
(\ref{decay}), a scenario called ``phase space
dominance''. But even this scenario requires stringent conditions on the dependence of 
cluster decay rates on the channel multiplicity (essentially like $A^N$) \cite{meaning} 
such that the exponential dependence of production rates on mass is not spoiled. 
Hence phase-space dominance is not less trivial in any way. A possible path to 
distinguish between the two scenarios is 
provided by the analysis of exclusive rates at low energy, although it must be pointed
out that the observed identical particle correlations already favors SHM which is
endowed with a built-in spacial extension, unlike phase-space dominance.

However, whether it is a proper thermal-statistical equilibrium in a finite volume or 
rather a phase space dominance effect, there must be a profound reason behind this 
phenomenon, which ultimately has to be related to the nature of QCD as a theory with 
strong coupling at large distances. Also, we believe that the intriguing
universality of the temperature found in elementary collisions 
as well as heavy ion collisions and its resemblance of the QCD critical 
temperature is not accidental.

If we assume that post-hadronization collisions are unable to restore equilibrium, how can a
quantum evolution process ensure it? Several years ago it was pointed out that a 
closed quantum system whose classical counterpart is chaotic and ergodic can give 
rise to thermal distributions provided that the so-called Berry conjecture applies 
\cite{srednicki}. Berry's conjecture \cite{berry} essentially states that the 
high-lying eigenfunction amplitudes $\psi({\bf x})$ in configuration space appear to be
random Gaussian numbers and, as a consequence, momentum space distribution is
microcanonical \cite{jarz}. This ``quantum thermalization'' mechanism has been invoked 
to explain the observed thermal-like distributions in hadronic processes \cite{andre}.
Of course, this argument requires that classical QCD is chaotic (as it has indeed been 
advocated \cite{biro}), that Berry's conjecture holds for quantum fields,
and that it can be applied to a dynamical process like hadronization.
All of these conditions are non-trivial, but pursuing these ideas further may give 
rise to interesting developments.

Recently, another appealing idea to explain the universality of thermal 
features in multihadron production has been put forward \cite{cks}. It invokes 
an analogy between confinement and black hole physics. It is conjectured that 
the phenomenon of confinement is equivalent to the formation of an event horizon 
for colored signals (quarks and gluons). Similarly to Hawking-Unruh radiation, 
the spectrum of hadrons, emitted as the result of a high energy collision, is 
thermal because no information can be conveyed from the causally disconnected 
region beyond event horizon. According to this so-called Hawking-Unruh scheme of 
hadronization, temperature is related to the string tension and is thereby universal. 
Another interesting consequence of this idea is that the extra strangeness suppression 
observed in elementary collisions can be quantitatively explained \cite{bcms} as an
effect of the different strange quark mass. 

These attempts relating the observed thermal features in hadron production processes
to quantum chaos or Hawking-Unruh radiation are still in a developmental stage.
Whether they will keep their promises will be seen in the future. Certainly, both 
share the vision that there is a fundamental quantum mechanical mechanism behind 
this phenomenon and no (or little) room for a classical collisional thermalization
process.

%********************************************************************************
\section{Quark Recombination} 
%********************************************************************************

The statistical hadronization model provides a successful description of key features 
of hadron production without explicitly invoking the underlying fundamental degrees 
of freedom in QCD, quarks and gluons. Clusters inherit the characteristics of partons 
from which they emerge, but it seems that there is no explicit role for
hadronic substructure in determining final hadron ratios except for the extra
strangeness suppression parameter $\gs$ which is a direct manifestation of the
differences between u, d and s quark masses. 
On the other hand, many hadronization models involve parton degrees of freedom explicitly
and it seems that some observables in relativistic heavy ion collisions 
at very high energy require the assumption of an underlying parton dynamics.

In a parton-based approach the goal is to calculate the probability
to produce a set of hadrons $h_1$, $\ldots$, $h_n$ from a given partonic ``initial'' 
state $\mathcal{C}$. In general, this problem involves interactions of partons at 
a scale around $\Lambda_{\mathrm{QCD}}$ which, as emphasized before, 
is a highly non-perturbative problem and as yet unsolved. In defining this problem, 
we have to overcome another obstacle which is connected to the preparation of the 
partonic ``initial'' state $\mathcal{C}$. Quarks and gluons are not asymptotic 
states. Rather, they usually appear as intermediate states in the scattering reaction 
$A+B \to \mathcal{C} +X \to h_1 + \ldots h_n +  X'$. Hence the partonic
state $\mathcal{C}$ is difficult to ``prepare'', and in fact the nature 
of quantum field theory requires us to integrate over all possible
states $\mathcal{C}$. In the worst case,
$\mathcal{C}$ might not be well defined at all because the partonic state
couples to the other states, $A$, $B$, $X$, etc. in a non-factorizable way.
In fact, this will always be true for any state $\mathcal{C}$ which is not
a color singlet, since the final hadron ensemble $h_1$, $\ldots$, $h_n$
is colorless and color must be exchanged with other parts of the system
during hadronization. We call this problem the factorization 
problem.

Several factorization schemes for hadronization have been developed starting
from the 1970s, resulting in some successful effective descriptions of 
hadronization. They are usually based on a proof that in a well defined
process a simple set of possible intermediate parton states $\mathcal{C}_i$ 
is giving the leading contribution to hadron production in powers of a small 
expansion parameter. The best known example is single hadron 
production in \ee or hadronic collisions at a large momentum scale 
$Q$. The leading contribution to the cross section in terms of powers of
$1/Q$ comes from intermediate states with 
just one single parton $\mathcal{C}_1=\{g\}$, $\mathcal{C}_2=\{u\}$, 
$\ldots$ and the cross section $\sigma^h$ can be factorized into a cross 
section for producing the intermediate parton $\mathcal{C}_i$ and a 
probability $D_{i/h}$ to produce $h$ from $C_i$ 
\cite{Owens:1986mp,Collins:1989gx,Albino:2008gy}
\begin{equation}
   \sigma^h \sim \sum_i  \sigma^{\mathcal{C}_i} \otimes D_{i/h} \, .
\end{equation}
Note that the leading contribution does not involve an interference 
effect between different partons in the 
amplitude and the complex conjugated amplitude and the process $C_i \to h$ 
can be formulated as a probabilistic problem with a probability distribution
$D_{i/h}(z)$ called a fragmentation function. $z$ is the fraction of the original
parton momentum carried by the hadron $h$.
In general, the functions $D_{i/h}(z)$ can not be calculated from first 
principles (which would be equivalent to fully solving the hadronization 
problem). However,
it makes single hadron production at large momentum treatable by separating
off the hadronization probability for a given hadron which can be measured
by experiment and is universally applicable to all processes where such a
factorization holds.

All hadronization models involving partons have to start from an assumption about a 
well defined parton ``initial state''. The exact details will depend on the process 
under investigation. Note that the SHM had such a basic axiom as well, defining
the probability of multihadron states in a cluster.
In this section, we discuss the model of quark recombination or coalescence. It also
involves a plausible but ad hoc assumption at the outset. This is less rigorous than
the factorization argument that can be given for parton fragmentation, but
it is phenomenologically very successful. We will focus on applications of quark 
recombination to heavy ion collisions.

%************************************************************
\subsection{Parton Fragmentation and Its Limitations}
\label{sec:frag}
%************************************************************

For a better understanding of recombination it is instructive to
investigate situations where parton fragmentation is failing.

Factorization of fragmentation functions off a hard parton scattering
cross section was first introduced for simple single scale processes 
like single-inclusive hadron productions \cite{Collins:1981uw}. Later 
the LEP Collider at CERN brought reliable data on hadron production in 
\ee collisions which allowed the extraction of universal sets of fragmentation 
functions \cite{Kniehl:2000fe,Albino:2005me,de Florian:2007hc} (see 
\cite{Albino:2008gy} for a modern review). In a physical gauge fragmentation 
functions have a straightforward definition in terms of counting operators 
$a_h^\dagger(P) a_h(P)$ of asymptotic hadron momentum eigenstates applied to the 
parton field.
E.g., if parton $i$ is a quark described by a field $\psi$ then 
we have
\begin{equation}
  D_{i/h}(z) \sim \int dx e^{-iPx/z} 
  \langle 0| \psi(0) a_h^\dagger(P) a_h(P) \bar\psi(x)|0\rangle
\end{equation}
where $x$ is a light cone coordinate conjugate to the light cone momentum 
$P$.

There are many situations where factorization involving a vacuum fragmentation
function could not be proven or is explicitly violated. In some situations
it is easy to see qualitatively why this is the case. One example
is hadron production at very forward rapidity in hadronic
collisions. A phenomenon known as the leading particle effect 
\cite{Adamovich:1993kc} can be observed if relative abundances of hadrons 
at forward rapidity are considered.
The FNAL E791 collaboration \cite{Aitala:1996hf} found
a large asymmetry between $D^-$ and $D^+$ mesons in fixed target 
experiments with $\pi^-$ beams on nuclei. Obviously a fragmentation
from $c$ or $\bar c$ quarks or from gluons should be nearly flavor blind
and the charm quark mass sets a (semi-)hard scale that can be used to justify
perturbative arguments. However, in this case the asymmetry comes
from recombination of the $\bar c$ from a $c\bar{c}$ pair produced in 
the collision with a $d$ valence quark from the beam $\pi^-$ remnants. 
This mechanism is enhanced compared to the $c$+$\bar d$ 
recombination which involves only a sea quark from the $\pi^-$
\cite{Braaten:2002yt}. Obviously the presence of the beam fragments acting
as spectators destroy the favorable conditions for vacuum fragmentation. 

The leading particle effect also motivated the birth
of the coalescence picture. Das and Hwa developed a model for
recombination of partons streaming forward in hadronic collisions 
\cite{Das:1977cp}.
There is no thermalized parton phase in this case, but this simple
model has many of the features of the recombination models for RHIC
we discuss below:
\begin{enumerate}
\item[(1)] The input is a (multi-)parton distribution $f(p_1,\ldots,p_n)$ 
  whose spectrum and chemical composition remain unchanged during the 
  hadronization process.
\item[(2)] Partons coalesce into hadrons according to recombination functions
  $\Phi$ which play the role of squared hadronic wave functions.
\item[(3)] Only valence quarks of the hadrons play a role in the hadronization 
  process.
\end{enumerate}
The number of recombining hadrons is therefore given by
\begin{equation}
  \label{eq:21}
  N_h \sim \Phi \otimes f(p_1,\ldots, p_n)
\end{equation}
where $n=2$ or 3 for mesons and baryons respectively.
The first condition is usually reinterpreted to mean that the 
coalescence process is fast enough such that the momentum distribution
and chemical composition of the partons do not react to the depletion
of partons during recombination. This \emph{assumption} addresses the question
of a well-defined ``initial'' parton state indicated in the last section.
The third condition has been the center of many debates. Where are the gluons
and sea quarks? While there is no calculation of this process from first
principles, we do have some qualitative arguments. First, the
momentum transfer involved is rather small and it is questionable whether
perturbative partons can be resolved. Second, we know that even in processes
that start at large momentum scales, degrees of freedom get frozen and give 
constituent masses to quarks \cite{Bowman:2002bm}. We could therefore 
argue that a dressing of quarks happens before the final recombination step. 
Third, we have no reason 
to believe that the hadrons originally formed are indeed free hadron states. 
There will necessarily be a formation time before we can actually interpret 
them as asymptotic states. 
The last two arguments are related to the role of chiral symmetry during 
hadronization. Chiral symmetry breaking has not been explicitly incorporated 
in recombination models thus far. 

%--------------------------------------------------------------------
\subsection{The Recombination Formalism}
%--------------------------------------------------------------------

\subsubsection{Basic Theory} 

Coalescence or recombination of particles appears in a wide array of systems
in atomic, molecular and plasma physics. As a first approximation
the details of the dynamical process are usually ignored. Rather, 
the adiabatic approximation of a projection of the initial 
multiparticle state onto the final coalesced state is considered. This 
instantaneous approximation is widely used in the literature for 
the case of partons coalescing into hadrons. The formalism introduced
here is also related to the successful coalescence model for nucleons
\cite{Kapusta:1980zz,coal-hadr}. After first applications to forward
hadron production and the leading particle effect the coalescence
concept was soon applied to heavy ion collisions 
\cite{Gupt:1983rq,Biro:1994mp,alcor,Zimanyi:1999py}.

The number of hadrons $h$ coalescing from a partonic system characterized by 
a density matrix $\hat\rho$ is given by
\begin{equation}
  \label{eq:0}
  N_h = \int\frac{d^3 P}{(2\pi)^3} \left\langle h;\mathbf{P}\right| \hat\rho 
  \left| h ;\mathbf{P}\right\rangle \, .
\end{equation}
Instantaneous here means that the states are defined on a hypersurface
which is typically either taken to be at constant time, $t=$ const.,
or on the light-cone $t=\pm z$. Note that the notion of instantaneous 
recombination guarantees that condition (1) from section \ref{sec:frag} is 
automatically fulfilled.

The instantaneous projection formalism has the conceptual disadvantage that 
only three components of the four-momentum are conserved in the
underlying $2\to 1$ or $3\to 1$ coalescence process. Qualitatively this
can be corrected by assuming that the participants can scatter off the
surrounding particles which can build up or dissipate their virtuality. 
At least in a 
\emph{equilibrium} state this should not change the momentum distribution 
or chemical composition of the particles. However, no explicit formalism 
has been developed to include this effect quantitatively. 

More dynamical approaches beyond the instantaneous projection approximation
have been considered as well \cite{Ravagli:2007xx}. They conserve 4-momentum 
by allowing a finite width for hadrons and will be discussed later. 
We first focus on instantaneous projection models that were discussed
for heavy ion collisions first by Greco, Ko and L\'evai [GKL] 
\cite{Greco:2003xt,Greco:2003mm}; Fries, M\"uller, Nonaka and Bass [FMNB] 
\cite{Fries:2003vb,Fries:2003kq,Fries:2003rf}; Hwa and Yang [HY] 
\cite{Hwa:2002tu,Hwa:2004ng} and Rapp and Shuryak [RS] \cite{Rapp:2003wn}.

>From Eq.\ (\ref{eq:21}) we can derive an expression for the number of mesons 
with a certain momentum $\mathbf{P}$ from recombination \cite{Fries:2003kq}
\begin{equation}
  \label{eq:22}
  \frac{dN_M}{d^3P} = \sum_{a,b} \int \frac{d^3 R}{(2\pi)^3} \frac{d^3q d^3r
  }{(2\pi)^3} W_{ab}\left( \mathbf{R}-\frac{\mathbf{r}}{2},
  \frac{\mathbf{P}}{2}-\mathbf{q};\mathbf{R}+\frac{\mathbf{r}}{2},
  \frac{\mathbf{P}}{2}+\mathbf{q} \right) \Phi_M (\mathbf{r},\mathbf{q}).
\end{equation}
Here $M$ denotes the meson and $a$, $b$ are its coalescing valence partons. 
$W_{ab}$ and $\Phi_M$ are the Wigner functions of the partons and the meson
respectively, $\mathbf{P}$ and $\mathbf{R}$ are the momentum and spatial 
coordinate of the meson, and $\mathbf{q}$ and $\mathbf{r}$ are related 
to the relative momentum and position of the quarks. The sum runs over 
all possible combinations of quantum numbers of valence quarks in the 
meson, which is usually replaced by a degeneracy factor $C_M$.
Eq.\ (\ref{eq:22}) is very intuitive for a hadronization hypersurface
at constant time. It can be made Lorentz-covariant to allow for relativistic
kinematics.

The corresponding formula for baryons, containing 3 valence quarks, is
completely analogous \cite{Fries:2003kq}. Note that Eq.\ (\ref{eq:22}) 
implements principle (3) from Section \ref{sec:frag} by taking into account 
only the lowest Fock state of the meson. It has been attempted to 
generalize Eq.\ (\ref{eq:22}) to include more partons which would be 
gluons or pairs of sea quarks, accounting for the next terms in a Fock 
expansion \cite{Muller:2005pv}. There is no difficulty in including 
these in the model, but data on elliptic flow scaling in heavy ion 
collisions puts bounds on the size of contributions from higher 
Fock states.

Let us recall the definition of the Wigner function for a meson consisting 
of two quarks
\begin{equation}
  \label{eq:2}
  \Phi_M (\mathbf{r},\mathbf{q}) = \int d^3 s e^{-i\mathbf{s}\cdot\mathbf{q}}
    \varphi_M \left(\mathbf{r}+\frac{\mathbf{s}}{2}\right)
    \varphi_M^* \left(\mathbf{r}-\frac{\mathbf{s}}{2}\right)
\end{equation}
where the 2-quark meson wave function in position space $\varphi_M$ can
be represented as
\begin{equation}
  \label{eq:3}
    \left\langle \mathbf{r}_1 ; \mathbf{r}_2 \big| M ; \mathbf{P} 
    \right\rangle = e^{-i\mathbf{P}\cdot (\mathbf{r}_1+\mathbf{r}_2)/2}
    \varphi_M \left( \mathbf{r}_1 - \mathbf{r}_2 \right )
\end{equation}
The Wigner function of the partons can be defined in a similar way
from the density matrix $\hat\rho$ \cite{Fries:2003kq}.
For further evaluation the Wigner functions have to be specified. Usually, 
the parton Wigner function $W_{ab}$ is replaced by its
classical limit, the multi-parton phase space distribution 
$f(x_1,p_1; \ldots; x_n,p_n)$ on the hypersurface of hadronization. The 
Wigner functions of the hadrons are not well known in this particular case. 
In general the wave function of a hadron depends in a non-trivial 
way on the frame of reference, the particular process, and the resolution 
scale. In heavy ion collisions the resolution scale is non-perturbative,
and of the order of the temperature at hadronization, $T_c< 200$ MeV.
Therefore the hadron Wigner functions are often modeled ad hoc. Luckily
the dependence of observables in heavy ion collisions on the shape of the 
wave function is suppressed as we will discuss below.

The different groups mentioned above using instantaneous recombination
mainly differ in details of how they apply the basic equation (\ref{eq:22})
to nucleus-nucleus and hadron-nucleus collisions \cite{Fries:2004ej}.
We will discuss some of their unique features and refer the reader to
the original literature and reviews \cite{Fries:2004ej,Fries:2008hs}
for further details.

\subsubsection{Different Implementations of Instantaneous Recombination} 

In the implementation by Greco, Ko and L\'evai [GKL] 
\cite{Greco:2003xt,Greco:2003mm} the full overlap integral in Eq.\ 
(\ref{eq:22}) over both relative position and momentum of the partons is 
calculated with Monte-Carlo techniques. In their case the Wigner function of the hadron 
is a simple product of spheres in position and momentum space with radii 
correlated through the uncertainty relation. 

In the implementations by Fries, M\"uller, Nonaka and Bass [FMNB] 
\cite{Fries:2003vb,Fries:2003kq,Fries:2003rf}; Hwa and Yang [HY] 
\cite{Hwa:2002tu,Hwa:2004ng} and Rapp and Shuryak [RS] \cite{Rapp:2003wn}
Eq.\ (\ref{eq:22}) is simplified by integrating out the information about 
position space. This leads to a pure momentum space formalism 
in which the information about the hadron is further compressed 
into a squared (momentum space) wave function. 
In a frame of reference in which one momentum component of the hadron is much
larger than the others the two momentum components transverse to the hadron
momentum can be integrated out as well, leading to a simple 1-dimensional 
overlap integral. 
Such a formalism is appropriate, e.g., for hadron production at large
transverse momentum in the laboratory frame.
It is also equivalent to the original formulation by Das and Hwa using 
recombination functions \cite{Das:1977cp}. 
The yield of mesons with momentum $P$ can then be expressed as
\begin{equation}
  \label{eq:6}
  \frac{dN_M}{d^3P} = C_M \int_\Sigma \frac{d\mathbf{\sigma}
  \cdot\mathbf{P}}{(2\pi)^3} 
  \int_0^1 dx_1 dx_2 \Phi_M(x_1,x_2) W_{ab}(x_1\mathbf{P};x_2\mathbf{P})
\end{equation}
where $d\mathbf{\sigma}$ is the hypersurface of hadronization.
In many cases the emission integral over the hypersurface 
is not calculated explicitly, but replaced by a normalization factor 
proportional to the volume of the hadronization hypersurface.
$x_1$ and $x_2$ are the fractions of momentum carried by the two valence
quarks.

For the lowest Fock state of a meson the squared wave function is mostly 
parameterized as \cite{Fries:2003kq}
\begin{equation}
  \label{eq:5}
  \Phi_M (x_1,x_2) = B x_1^{\alpha_1} x_2^{\alpha_2} \delta(x_1+x_2-1) \, .
\end{equation}
The powers $\alpha_i$ determine the shape while the constant
$B$ normalizes the integral over $\Phi_M$ to unity.
Light cone distribution amplitudes suggest $\alpha_i=2$ for a light quark 
system \cite{Chernyak:1984bm,Aitala:2000hb,Bakulev:2001pa}.
Values $\alpha_c =5$, $\alpha_{u,d}=1$ are used for the charm-light quark 
system in $D$ mesons in Ref.\ \cite{Rapp:2003wn}.
The extreme case of the two quarks exactly sharing the momentum,
 $\Phi_M (x_1,x_2) = \delta(x_1-1/2)\delta(x_2-1/2)$, is often
considered for schematic estimates. Formally this corresponds to
$\alpha_i \to \infty$ with the ratio of the $\alpha_i$ fixed. 

\subsubsection{Dynamical Recombination}

One of the drawbacks of the instantaneous approximation is the
violation of energy conservation which has to be restored, on
average, by interactions in the medium which are not further
specified. One early attempt to improve the situation was the
introduction of a mass distribution for the quarks \cite{Zimanyi:2005nn}.
The mass distribution can be thought of as an effective way to incorporate 
some in-medium effects.
This approach allows to enforce both momentum and energy conservation 
and one finds fairly good agreement with data for transverse momentum
spectra in heavy ion collisions.

A promising new direction has recently been taken by Ravagli and Rapp 
(RR) \cite{Ravagli:2007xx,Ravagli:2008rt}. They advocate a dynamical 
approach in which the instantaneous projection of quark states onto hadron 
states is
replaced by a transport picture which involves ensembles $f_a$ and 
$f_M$ of quarks and mesons evolving with time. To determine
properties of the meson distribution starting from given quark distributions
one solves the Boltzmann equation 
\begin{multline}
  \frac{\partial}{\partial t} f_M(t,P) = -\frac{\Gamma}{\gamma_P} f_M(t,P)
  + \int\frac{d^3p_1 d^3 p_2}{(2\pi)^6} \int d^3 x 
  f_a (x,p_1) f_b (x,p_2) \\ \times \sigma(s)
  v(p_1,p_2) \delta^{(3)}(\mathbf{P}-\mathbf{p}_1-\mathbf{p}_2)
\end{multline}
with a gain term describing the scattering of a quark-antiquark pair $a$-$b$ 
into a meson state $M$ with finite width $\Gamma$ and a corresponding
loss term describing the decay of $M$ into a pair $a$-$b$. Here $\sigma(s)$ 
is the resonant cross section in Breit-Wigner form with width $\Gamma$, $v$ 
is the relative velocity of the pair and $\gamma_P=\sqrt{\mathbf{P}^2-m^2}/m$ 
is the gamma factor of the meson with mass $m$.

This implementation naturally conserves 4-momentum and permits, e.g., the 
study of kinetic energy scaling \cite{Afanasiev:2007tv,Ravagli:2008rt}. 
The problem of baryon formation 
is slightly more difficult in a transport approach and has not yet been
addressed. We should also note that even for a finite meson width the 
creation of light Goldstone bosons, in particular pions, is still
problematic. They are much lighter than the sum of the participating quark 
masses which are usually taken to be constituent quark masses.

%-------------------------------------------------------------------
\subsection{Connection with other Hadronization Models}
%-------------------------------------------------------------------

The recombination model was conceived as a correction to vacuum fragmentation
in order to include multi-parton effects. Indeed, for a first, intuitive understanding 
it is helpful to view parton fragmentation and parton recombination as two 
limiting cases of hadronization. The former fixes the parton state
$\mathcal C$ to be a single parton in the vacuum. The latter assumes the
parton state to be a subset of a populated, maybe even thermalized
phase space, such that valence partons are readily available. This
simple picture allows a basic understanding of most qualitative features
of both mechanisms.

Recombination is the most effective way to form hadrons from a 
thermalized parton phase space, at least for not too small momentum 
$P$. This is easy to see in the instantaneous momentum
space formalism, but it can be shown to hold numerically in more general 
cases as well. Let us assume that the partons at moderate to large values 
of $p$ (large enough that the energy $E$ can be approximated by 
$E\approx p$)  are distributed according to a Boltzmann distribution 
$f_a = e^{-p/T}$. 
Fragmentation then leads to a hadron spectrum which has an effective slope
$\langle z \rangle T$ with $\langle z \rangle < 1$ being the mean value
of the momentum fraction $z$ of the hadron from a fragmentation function
$D(z)$. Hence the resulting slope is steeper than $1/T$.
On the other hand, hadrons recombining according to Eq.\ (\ref{eq:5}) 
have a distribution
\begin{equation}
  f_M \sim \int dx e^{-xP/T} e^{-(1-x)P/T} \Phi(x) \sim e^{-P/T}
\end{equation}
for mesons, and similar for baryons. Hence we conclude that the recombination 
formalism is indeed more effective for large $P$ as long as the spectrum is 
exponential.

We also notice that for large momenta ($P >> M$) we recover the results
we obtain from the (momentum dependent) statistical model. At least this
is true if the same set of resonances is included. In practice, this has not yet
been implemented for recombination. Only a few studies have 
been dealing with resonances in a limited way 
\cite{Greco:2007nu,Greco:2003mm,Nonaka:2003ew}. Of course, 
one can argue that the contributions from resonance decay to stable hadron 
spectra are formally vanishing in the limit $P\to \infty$. With this caveat
the correspondence between recombination and statistical hadronization
would be precise.

Unfortunately, the limit $P\to \infty$ is an academic one, since in any
realistic system the parton spectrum will eventually become a power 
law $f_a \sim p^{-\alpha}$ for increasing momentum $p$. Once this kinematic
regime is reached fragmentation becomes more effective, producing hadrons
with a slightly shifted power law $\sim P^{-\alpha-\delta}$, where the small
exponent $\delta$ approximates the changed slope after applying the 
fragmentation function. On the other hand the power is doubled or tripled 
for recombining mesons ($\sim P^{-2\alpha}$) or  baryons ($\sim
P^{-3\alpha}$), respectively. Hence, in any realistic system 
fragmentation dominates at asymptotically large momentum, in accordance 
with QCD factorization theorems. 

One might ask whether recombination is interesting to consider given its
apparently small region of applicability.
Curiously, in heavy ion collisions at RHIC we find a kinematic region 
at intermediate transverse momentum between 1.5 and 6 GeV/$c$ where 
recombination still dominates over fragmentation, and the momenta are 
large enough compared to the masses involved such that the simple
instantaneous projection formalism, even without including resonances, 
gives very reasonable results for hadron spectra and elliptic flow. 
Moreover, unlike statistical hadronization, recombination can make certain
predictions connecting the kinematic behavior of hadrons and quarks.
Elliptic flow scaling in heavy ion collisions is a famous example. It must 
be connected to hadronization because it exhibits a dependence on the number 
of valence quarks. We will offer a survey of recent experimental results
supporting quark recombination further below.

Let us summarize this subsection so far. We have found that there is a 
formal connection between parton recombination and the statistical
hadronization model at large momenta. Without doubt a
successful implementation of parton recombination at low momenta 
in heavy ion collisions should reproduce the statistical model in that
kinematic region as well. Such a description is not available yet. It would
have to include energy and momentum conservation, at least in a statistical
sense, as well as the full range of resonances and their decays.
This issue of resonances is also related to entropy conservation. Direct 
coalescence into stable hadrons seems to reduce the number of particles by 
about a factor two, raising questions about the second law of
thermodynamics. At intermediate and large momenta, where 
the current formalism is applicable, only a minute fraction of 
the total number of particles is residing, so that entropy is not of 
any grave concern here.

We conclude this subsection by discussing some further connections
between recombination and fragmentation.
We return to the original question of preparing a well-defined partonic
state $\mathcal{C}$. The fragmentation formalism puts a cut between
long- and short-distance physics which is well motivated because we can not
calculate the former perturbatively, while we can compute the latter. 
The single parton state is treated like an observed final state of the 
perturbative calculation. However we know that many processes 
during fragmentation are still partonic in nature, not hadronic.
Hence we are tempted to ask whether we could define a partonic state
$\mathcal{C'}$ later in the development of the jet parton shower.
A jet shower might include a dozen or more partons which at some point
feel the pressure from the QCD vacuum to bind into hadrons. This binding
might well be described again by recombination.

It is not at all clear that such a parton ensemble in a jet ``just before''
hadronization can always be well defined. However, the cluster picture of
statistical hadronization seems to support this idea. One can try to choose
this as a starting assumption. This picture of the 
fragmentation process has been implemented in the recombination model of 
Hwa and Yang \cite{Hwa:2003ic}.  Instead of fragmenting hard partons 
directly, they define the parton shower of a jet (initiated by a hard 
parton) through so-called shower distributions. These are given by 
non-perturbative splitting functions $S_{i/j}(z)$ which describe the 
probability to find a parton of flavor $j$ with momentum fraction $z$ 
in a jet originating from a hard parton $i$. The parton content of a 
single jet is then allowed to recombine and should match the 
corresponding fragmentation functions. E.g. for a meson $M$ with valence
quarks $a$ and $b$
\begin{equation}
  D_{i/M}(z) = \int dx_a dx_b \left[ S_{i/a}(x_a) 
  S_{i/b}\left(\frac{x_b}{1-x_a}\right) + (a\leftrightarrow b) \right] 
  \Phi_M(x_a/z,x_b/z) \, .
\end{equation}
The shower distributions are not determined from
first principles but are fitted to reproduce the known fragmentation functions 
for pions, protons and kaons \cite{Hwa:2003ic}. 

The power of this approach lies in the fact that the fragmentation 
part of the hadron spectrum is computed with the same recombination
formalism that is applied to the thermalized part of phase space.
It is then very natural to allow coalescence of shower partons with thermal
or soft partons \cite{Hwa:2004ng} by defining a total parton distribution
\begin{equation}
  f_a(p) = f_a^{\mathrm{soft}}(p) + f_a^{\mathrm{shower}}(p) =: T(p) + S(p).
  \label{eq:hardsoft}
\end{equation}
which is subject to the recombination formalism. For mesons we obtain
two known contributions. The term $\sim SS$ corresponds to the coalescence
of two partons from the same jet and reproduces fragmentation, the term
$\sim TT$ recombines two soft partons and reproduces the soft recombination
spectrum discussed in the previous subsection. The mixed term $\sim TS$
is new and describes the pick-up of a soft parton by a jet parton to form a 
meson. For baryons two of these mixed terms, $\sim TTS$ and $\sim TSS$, exist.
These soft-hard coalescence terms make important contributions in some cases. 
They can, e.g., explain the final state dependence of the Cronin effect 
observed in $d+A$ collisions at RHIC \cite{Hwa:2004zd,Hwa:2004in}, 
and lead to jet-like correlations between hadrons
\cite{Hwa:2004sw,Fries:2004hd,Fries:2004gw,Fries:2005is}.

To summarize, there is reason to believe that parton coalescence could be
an overarching qualitative concept for hadronization, based on its
phenomenological success. However, so far it is only defined 
in a naive parton model sense. Beyond that it is not clear that the parton 
content of an arbitrary scattering of hadrons or nuclei before 
hadronization can be written down in a well-defined way. E.g.\ in Eq.\ 
(\ref{eq:hardsoft}), it is not clear why hadrons forming at very 
different times in the soft and hard sector can be treated together. It is
also not completely understood how the scale dependence of the
fragmentation functions can be interpreted correctly
\cite{Majumder:2005jy}.

%--------------------------------------------------------------------
\subsection{Experimental Evidence for Recombination}
%--------------------------------------------------------------------

We have already discussed the leading particle effect and mentioned
the Cronin effect in hadron-nucleus collisions. Here, we want to
review some of the support for recombination from heavy ion collisions.
Three early results from the Relativistic Heavy Ion Collider led to the
conclusion that hadrons at intermediate transverse momentum 
$p_T$ (1.5~GeV/$c$ $< p_T < 4\ldots 6$~GeV/$c$), are produced by recombination
of quarks.
\begin{itemize} 
\item[(a)] The baryon-to-meson ratios in central Au-Au collisions was found 
  to be enhanced~\cite{Adler:2003kg,Adams:2006wk}.
  The proton/pion ratio $\approx 1$ is incompatible with expectations from 
  fragmentation.
\item[(b)] The nuclear modification factors $R_{AA}$ and $R_{CP}$,
  i.e.\ the ratio of yields in central Au-Au collisions compared to 
  pp and peripheral collisions, resp., scaled by the number of binary 
  nucleon-nucleon collisions, is not suppressed for 
  baryons~\cite{Adler:2003kg,Adams:2003am}. 
\item[(c)] The anisotropy of particle production in azimuth 
  relative to the reaction plane, called the elliptic flow parameter 
  $v_2$, scales in a universal way for mesons and baryons 
  \cite{Adams:2003am,Adams:2004bi,Adler:2003kt,Adams:2005zg}.
\end{itemize}

\subsubsection{Hadron Spectra and Ratios}

\begin{figure}[tb]%3	% Figure using psFig.sty 
\centerline{\epsfig{file=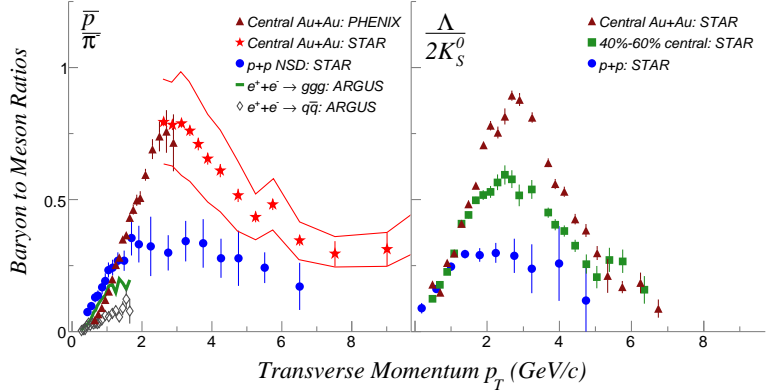,width=25pc}} 
\caption{ Left: $\overline{p}/\pi^{-}$ ratios measured in central Au-Au
  collisions at $\sqrt{s_{_{NN}}}=200$~GeV at RHIC, compared to
  measurements from \ee and pp collisions. Right: The ratio
  $\overline{\Lambda}$/2$K^0_S$ for central and mid-central Au-Au
  collisions at $\sqrt{s_{_{NN}}}=200$~GeV measured by STAR. The
  $\overline{p}/\pi^-$ ratio from pp collisions from STAR is shown
  for comparison.}
\label{B/M} 
\end{figure}

Fig.~\ref{B/M} shows the measured anti-proton/pion~\cite{Adler:2003kg}
and $\overline{\Lambda}$/$K_S^0$~\cite{Adams:2006wk} ratios as functions of
transverse momentum $p_T$ for various centralities and collision systems. 
At intermediate $p_T$ the baryon-to-meson ratios in central Au-Au collisions 
are significantly larger than those in \ee~\cite{Abreu:2000nw} or pp
collisions~\cite{Alper:1975jm}. These results were the first clear 
indication that hadronization proceeds differently in Au-Au collisions and
pp collisions even in a regime traditionally expected to be dominated by
jet fragmentation.

    \begin{figure}[bt]	% Figure using psFig.sty 
          \centerline{\epsfig{file=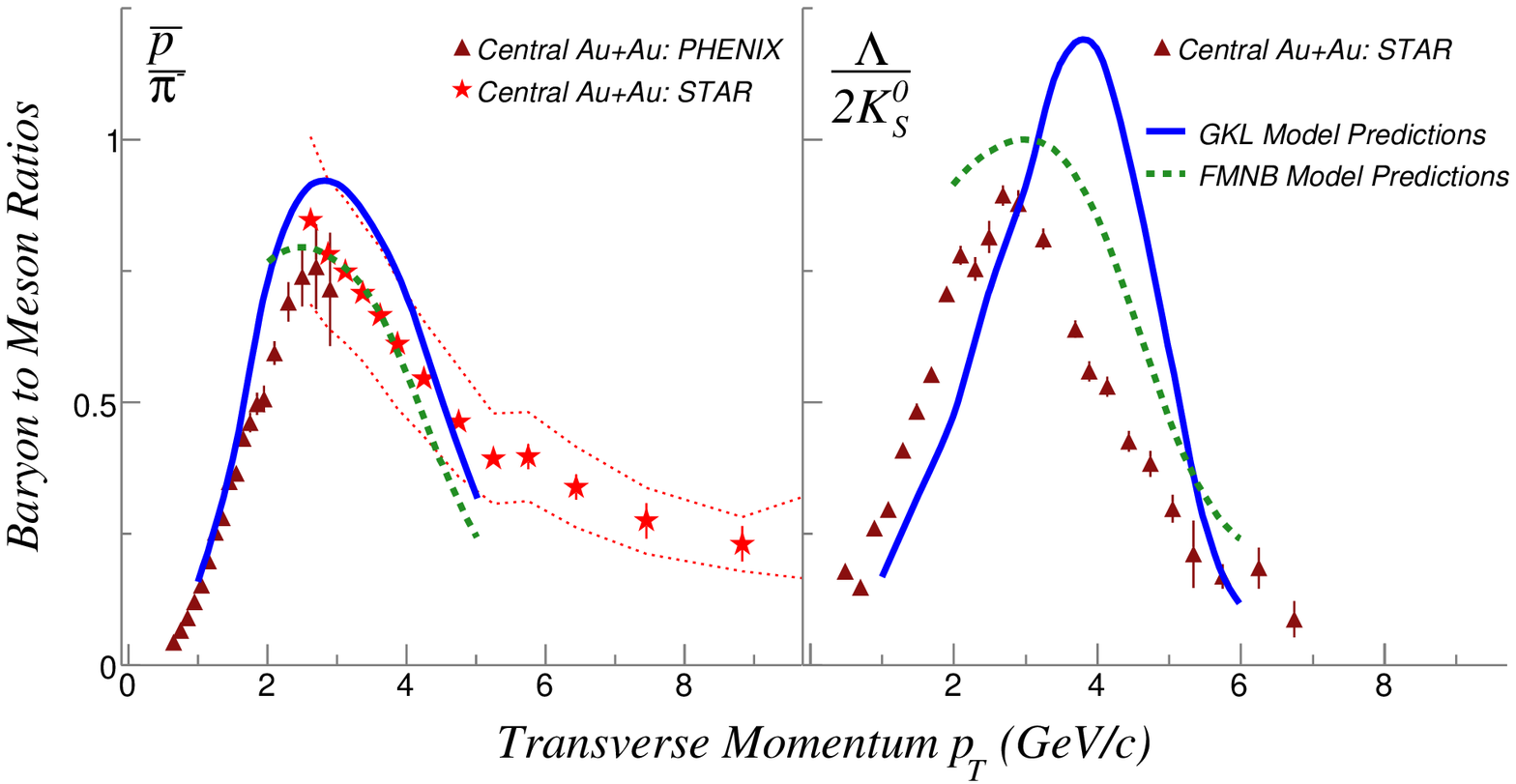,width=25pc}} 
          \caption{Ratios of baryon yields to meson yields for
            central Au-Au collisions at 200 GeV. The GKL and FMNB 
            calculations for $\bar p/\pi^-$ (left) and $\Lambda/2K_s^0$
            are compared to STAR and PHENIX data.}
          \label{ratios} 
        \end{figure}

Fig.\ \ref{ratios} shows anti-proton/pion (left) and 
$\overline{\Lambda}$/$K_S^0$ (right) ratios compared with instantaneous
coalescence calculations from Greco, Ko and L\'evai [GKL] 
\cite{Greco:2003mm,Greco:2005sn} and Fries, M\"uller, Nonaka and Bass [FMNB] 
\cite{Fries:2003kq,Nonaka:2003hx}.
Despite some deviations these calculations reproduce the characteristic
peak structure very well. This peak comes from the interplay of soft
production which leads to steeply rising ratios at low $p_T$, and low ratios at
large $p_T$ coming from jet fragmentation. Recombination gives an 
explanation why soft production dominates out to intermediate $p_T$.

\begin{figure}[bt]%3	% Figure using psFig.sty 
\centerline{\epsfig{file=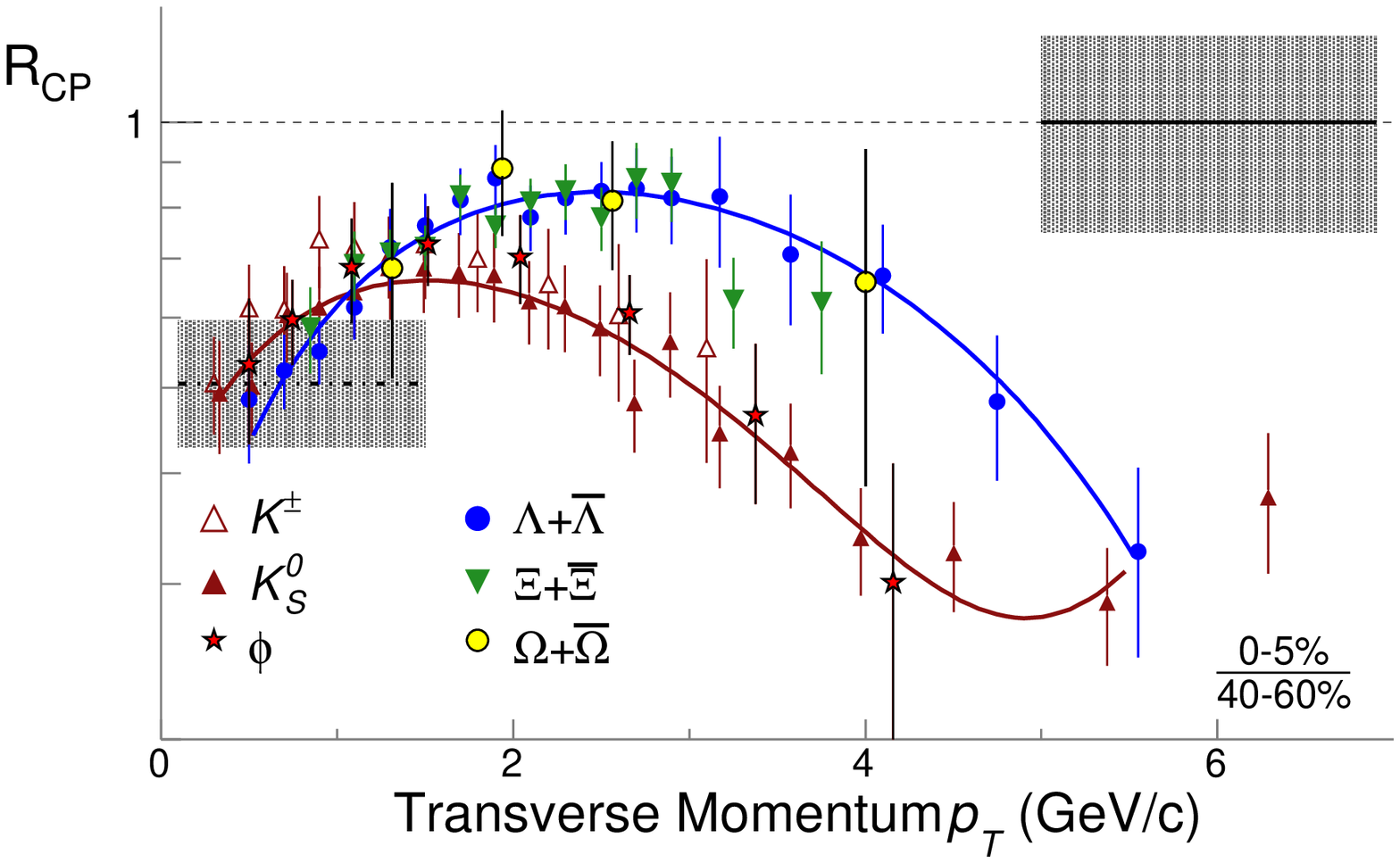,width=25pc}} 
\caption{Nuclear modification factors ($R_{CP}$) for various 
  identified particles measured in Au-Au collisions at 
  $\sqrt{s_{_{NN}}}=200$~GeV by the STAR collaboration.
  The values for baryons and mesons fall into two separate bands 
  (indicated by lines to guide the eye) with the baryon $R_{CP}$ 
  larger than the meson $R_{CP}$.} 
\label{rcp} 
\end{figure}

The different behavior of baryons and mesons is also expressed in different
nuclear suppression factors. Fig.\ \ref{rcp} shows the nuclear 
modification factor $R_{CP}$ measured at RHIC for various identified 
hadron species. If a nuclear collision is a simple superposition of binary 
nucleon collisions, $R_{CP}$ will equal one. Therefore the suppression of 
$R_{CP}$ at large $p_T$ is a signature for jet quenching \cite{dedx}. 
The baryons ($\Lambda+\overline{\Lambda}$, $\Xi+\overline{\Xi}$, and
$\Omega+\overline{\Omega}$) \cite{Adams:2006wk,Adams:2006ke}, 
show significantly less suppression than mesons (kaons or 
$\phi$) \cite{Adams:2004ep,Abelev:2007rw}). The same behavior was found 
for protons and pions \cite{Abelev:2006jr}.
The second important observation is that the behavior is universal for
all baryons and all meson separately, and largely independent of mass. 
Heavy $\phi$ mesons behave like pions, not like the equally heavy protons 
or $\Lambda$ baryons \cite{Abelev:2007rw,Afanasiev:2007tv}.
This is a clear indication that the decisive factor here is the
number of valence quarks.

        \begin{figure}[bt]	% Figure using psFig.sty 
%          \centerline{\epsfig{file=reco_spec.eps,width=25pc}}
          \centerline{\epsfig{file=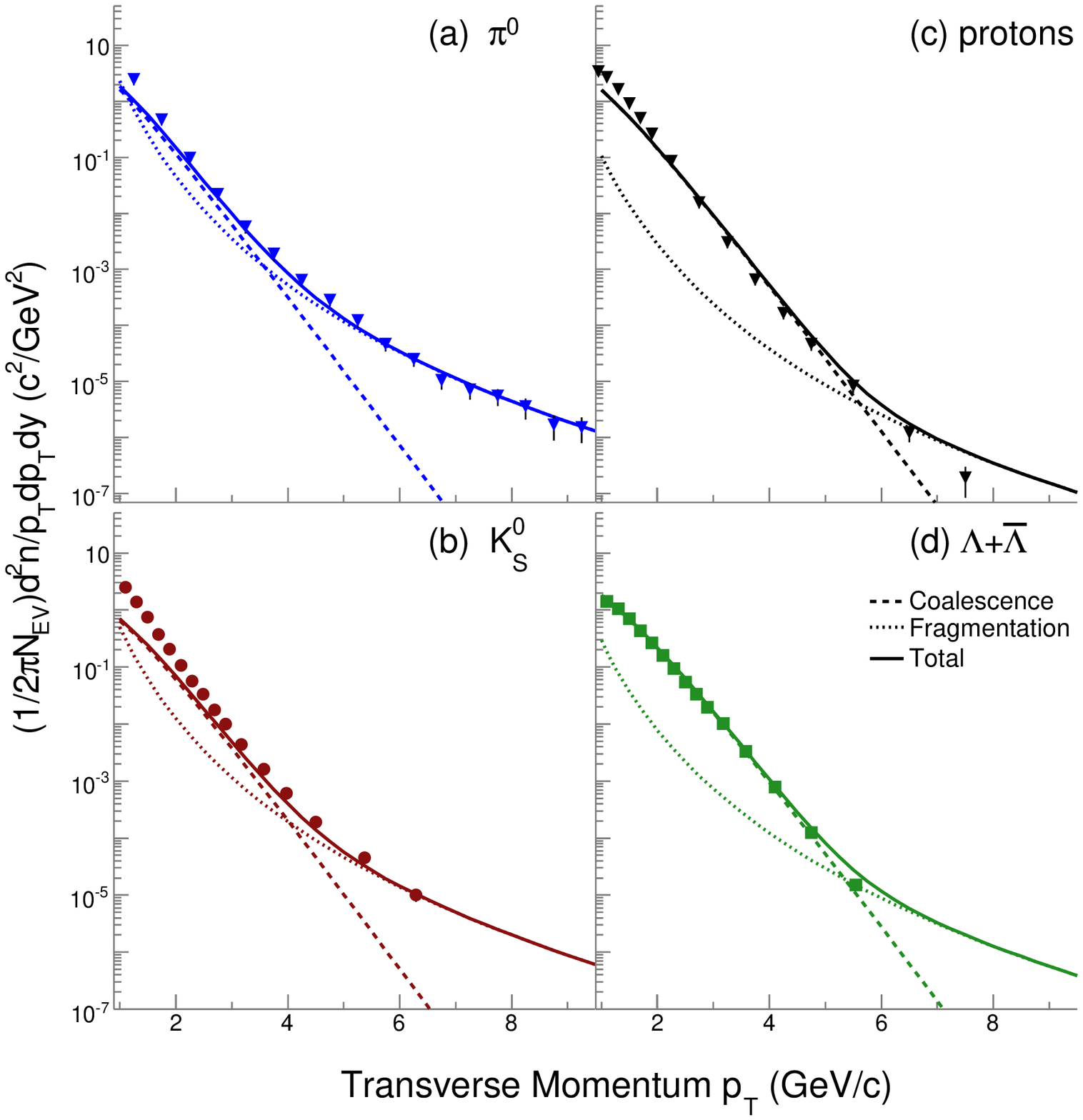,width=25pc}} 
          \caption{ Hadron $p_T$-spectra at midrapidity from 200 GeV
            central Au-Au collisions. The curves show the
            recombination and fragmentation components of the spectra
            obtained in the FMNB formalism along with the total which 
            compares well with the data.}
          \label{spec} 
        \end{figure}

Fig.\ \ref{spec} shows spectra from the FMNB coalescence model 
\cite{Fries:2003kq} for neutral pions, kaons, protons and hyperons in
central Au-Au collisions compared to data from RHIC
\cite{Adams:2006ke,Abelev:2006jr}. The recombination region is 
visible as an exponential slope at intermediate $p_T$ with a 
transition to a power-law shape at higher $p_T$. Note that 
the FMNB model simply adds recombination and fragmentation contributions.
Nevertheless the description of the data is quite good above 2 GeV/$c$.
Note that the transition from an exponential shape to a power-law shape 
happens at a higher $p_T$ for baryons than it does for mesons. This is
expected and due to the inherent suppression of baryons in vacuum 
fragmentation which makes coalescence competitive up to quite large $p_T$.

\subsubsection{Elliptic Flow}

Recombination makes a straightforward prediction for the elliptic
flow $v_2$ of hadrons \cite{flow_methods}, using the elliptic flow of 
quarks as input. This behavior can be easily derived in the 
instantaneous momentum-space formalism.
Let us assume that the elliptic flow of a set of partons $a$ just
before hadronization is given by an anisotropy $v_2^a(p_T)$ at
mid-rapidity. The phase space distribution of partons $a$ can
then be written in terms of the azimuthal angle $\phi$ as
\begin{equation}
  \label{eq:9}
  f_a(\mathbf{p}_T) = \bar f_a(p_T)\left( 1+ 2v_2^a(p_T) \cos2\phi
  \right) \, ,
\end{equation}
where odd harmonics are vanishing due to the symmetry of the system
and higher harmonics are neglected. $\bar f$ is the distribution
averaged over the azimuthal angle $\phi$.

 For a meson with two valence partons $a$ and $b$ and for small elliptic flow
$v_2 \ll 1$ one has \cite{Fries:2003kq}
\begin{eqnarray}
  \label{eq:10}
  v_2^M(p_T) &=& \frac{\int d\phi \cos(2\phi) dN_M/d^2p_T}{\int d\phi
    dN_M/d^2p_T}  \\
  && \sim \int dx_a dx_b \Phi_M(x_a,x_b) \left[ v_2^a(x_ap_T) + v_2^b(x_bp_T)
  \right]  \, .  \nonumber
\end{eqnarray}
In the case of a very narrow wave function in momentum space 
($\alpha\to \infty$) this leads to the expression
\begin{equation}
  \label{eq:11}
  v_2^M(p_T) = v_2^a(x_a p_T) + v_2^b(x_b p_T)\, .
\end{equation}
with fixed momentum fractions $x_a$ and $x_b$ ($x_a+x_b=1$).

Thus for hadrons consisting of light quarks which exhibit the same
elliptic flow before hadronization we arrive at a simple scaling law
with the number of valence quarks $n$:
 \begin{equation}
  \label{eq:12}
  v_2^h(p_T) = n v_2^a(p_T/n)  \, .
\end{equation}
This scaling law had first been found in data from RHIC and was
quickly interpreted as a coalescence artefact \cite{Voloshin:2002wa}.
Since then this connection has been solidified in almost all
coalescence models \cite{Fries:2003kq,Greco:2005jk,Molnar:2003ff}. 
Eq.\ (\ref{eq:11}) has also been used to extract the elliptic flow of 
heavy quarks from measurements of $D$ mesons \cite{Lin:2003jy,greco-c}.
The line of thought presented for elliptic flow can be extended to harmonics 
beyond the second order. Generalized scaling laws for the 4th and 6th order
harmonics have been derived in Ref.~\cite{Kolb:2004gi}.

\begin{figure}%3	% Figure using psFig.sty 
         \centerline{\epsfig{file=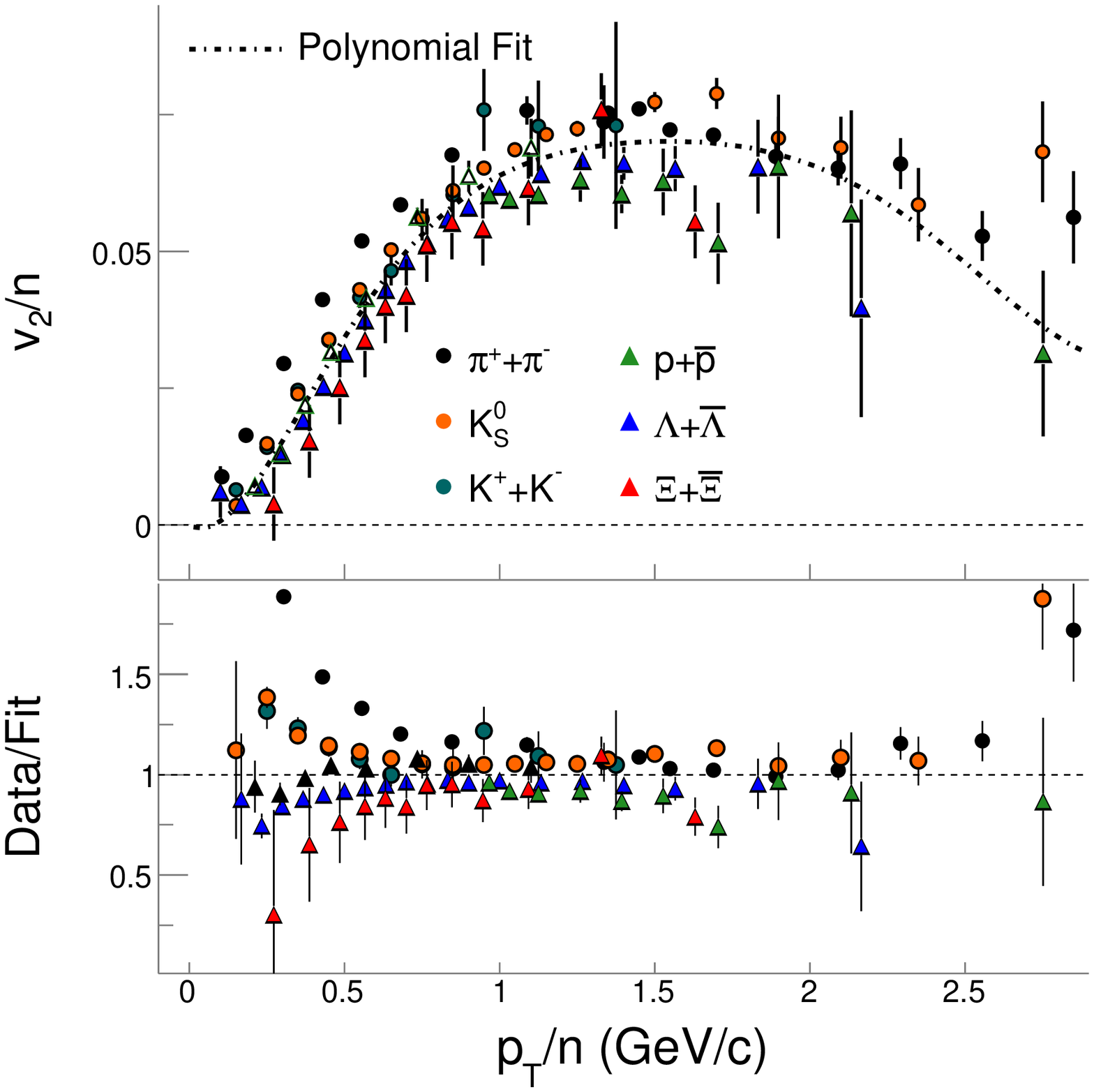,width=17pc}
                     \epsfig{file=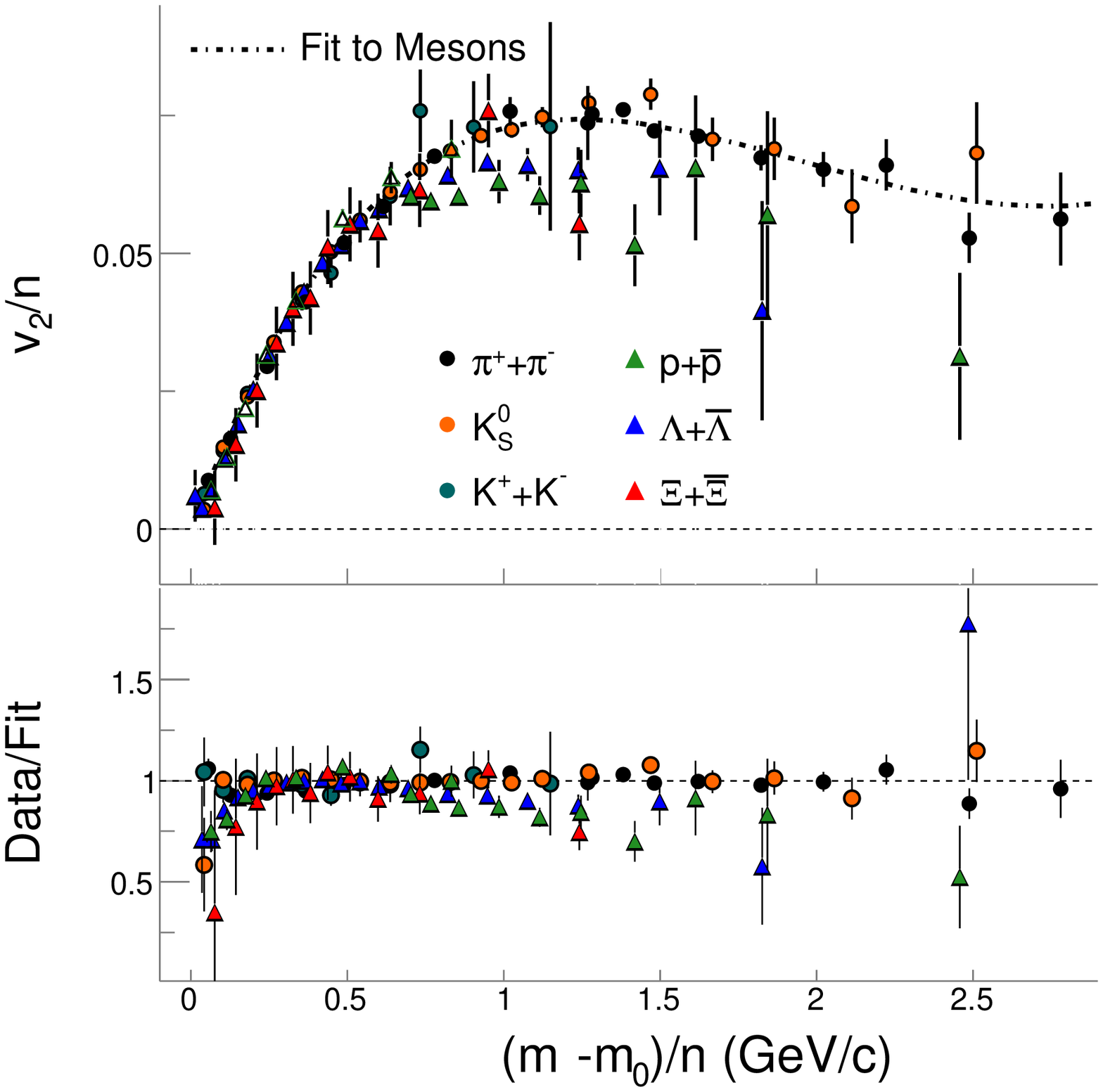,width=17pc}} 
          \caption{ Top left panel: The elliptic anisotropy parameter $v_2$
            scaled by quark number $n$ and plotted vs $p_T/n$. A
            polynomial curve is fit to all the data. The ratio of
            $v_2/n$ to the fit function is shown in the bottom left panel.
            Right panels: same but with $p_T$ replaced by $m_T-m_0$}
           \label{fig4} 
\end{figure} 

Fig.~\ref{fig4} shows data on $v_2$ scaled by the number $n$ of valence 
quarks in a given hadron as a function of $p_T/n$ for several species of
identified hadrons at $\sqrt{s_{_{NN}}}$ = 200 
GeV~\cite{Abelev:2007rw,Abelev:2008ed}.
To investigate the quality of agreement, the data from the top
panel are scaled by a polynomial fitted to the universal curve and plotted 
in the bottom panel. 
At low $p_T$ hydrodynamics \cite{hydro,Kolb:2003dz,Hirano:2003pw} predicts 
an ordering of $v_2$ by hadron mass which is confirmed by data 
\cite{Adams:2003am,Adler:2003kt,pidv2}.
That is the reason why valence quark scaling of $p_T$ is not working for 
$p_T/n < 0.5$ GeV/$c$. The two phenomena can be reconciled by first scaling 
the $v_2$ results in transverse kinetic energy $E_T= m_T-m_0$, and 
then applying a scaling in
$n$ \cite{Afanasiev:2007tv}. The result is also shown in Fig.\ \ref{fig4}.
The scaling at low $m_T-m_0$ holds to high accuracy.

The experimentally observed scaling is a spectacular success for the 
recombination picture. However, it is also the source for some open
questions. First, the derivation of the scaling law in the instantaneous
momentum space formalism is rather simplistic and neglects both energy
conservation and space-momentum correlations in the quark distributions.
The observed scaling seems to put tight constraints on the latter
\cite{Pratt:2004zq,Molnar:2004rr}. The instantaneous formalism is also
ill-prepared to distinguish between valence quark scaling in $p_T$ and $E_T$
because of the partial violation of energy and momentum conservation.
Recent work using dynamical recombination with energy conservation
and realistic space-momentum correlations found that recombination 
respects scaling in $E_T$ \cite{Ravagli:2008rt}. Future work in this 
direction is required to understand all aspects of the scaling 
phenomenon and its breaking \cite{Greco:2005jk,Lin:2003jy,
Muller:2005pv,decayv2}. It will also be interesting to see how prominent
recombination effects will be at the higher energies reached at the 
Large Hadron Collider LHC \cite{Fries:2003fr}.

%*********************************************************************
\section{Summary and conclusions}
%*********************************************************************

In this chapter, we have presented two models of hadronization
motivated by phenomenological observations in elementary and nuclear
collisions. 

The statistical model postulates that hadronization proceeds through the 
formation of massive colorless objects called clusters or fireballs, 
according to the general ideas of preconfinement of other cluster hadronization
models \cite{herwig}. They are formed in fragmentation in \ee
collisions just as they are in relativistic heavy ion collisions. While
the size and charge distribution of these clusters differ between collision 
systems, each cluster gives rise to multihadronic states in a purely statistical
fashion. Mathematically this can be described by a density matrix of
localized hadronic states projected onto eigenstates with all
possible hadronic quantum numbers. Interactions are, to a first approximation
accommodated by including hadronic resonances (ideal hadron-resonance gas model).

The statistical model has been applied to a wide variety of
small and large systems which differ by more than two orders of magnitude in the
collision center of mass energy $\sqrt{s}$, revealing intriguing universal 
features of the hadronization process. The assumptions of the model
were found to hold in a remarkable way for relative abundances and transverse
momentum spectra of both light and heavy flavored species.
For large enough ensembles of clusters a canonical statistical description
can be used, which allows the introduction of a temperature $T$ of the 
clusters which is typically between 160 and 170 MeV, corresponding to an 
average energy density of $\epsilon\approx 0.5$ GeV/fm$^3$ in a single
cluster. An additional strangeness suppression parameter $\gamma_s$ is 
needed to reproduce the data, which is found to be between 0.5 and 0.7 in 
elementary collisions, while $\gamma_s \approx 1$ in central heavy ion 
collisions at RHIC. The origin of such an additional parameter and its relation
to the strange quark mass are not clarified yet.

Clusters of hadrons at hadronization capture several important features
of hadron production. It remains unspecified in this model how to connect 
it to the previous partonic phase in a microscopic picture. This step is complicated
by the nature of the final partonic stage as an intermediate
state in a large quantum system. This problem has been successfully resolved
only for two simple cases: for single partons in high energy collisions 
described by vacuum fragmentation, and for the thermalized parton 
phase in heavy ion collisions which hadronizes through parton
coalescence. 

We have discussed the quark recombination model for heavy ion collisions
and we have shown that this model recovers important features of the 
statistical model if applied to a thermalized parton phase. Experimental
evidence for quark recombination comes from the Relativistic
Heavy Ion Collider, in particular the spectacular valence quark scaling
of elliptic flow. This is an important step to prove the existence of
collective effects on the level of partons in heavy ion collisions
\cite{Sorensen:2007rk}.
While instantaneous recombination cannot capture the bulk of the 
hadronization process and is only applicable for momenta larger than 1 or 2 
GeV/$c$, more recent models based on transport descriptions
are able to address the urgent questions of energy conservation
and entropy during hadronization.
We have also shown a possible way to reconcile parton fragmentation with
statistical hadronization via a recombination model for the full shower of
partons created inside a jet cone.

Quark recombination and the statistical hadronization model offer a
successful phenomenological description of many aspects of hadron
production in elementary and nuclear collisions. They are based on
few core principles which seem to capture key universal features in the process 
of hadron formation. While they do not answer the central questions of confinement 
and chiral symmetry breaking, which are of fundamental importance for 
QCD, they allow us to understand some of the key properties of QCD in the 
non-perturbative regime.

%**************************************************************************
\section*{Acknowledgements}
We are indebted to R.\ Stock for his encouragement and invaluable help
while preparing this review. We thank P.\ Sorensen for his permission to use 
Figs.\ \ref{B/M}-\ref{fig4}. 
%*************************************************************************

%%%%%%%%%%%%%%%%%%%%%%%% referenc.tex %%%%%%%%%%%%%%%%%%%%%%%%%%%%%%
% sample references
% %
% Use this file as a template for your own input.
%
%%%%%%%%%%%%%%%%%%%%%%%% Springer-Verlag %%%%%%%%%%%%%%%%%%%%%%%%%%
%
% BibTeX users please use
% \bibliographystyle{}
% \bibliography{}
%

\end{document}